\documentclass[useAMS,usenatbib]{mn2e}
\usepackage{graphicx}
\usepackage{amsmath}
\usepackage{txfonts}
\usepackage{rotating}
\usepackage{lscape}
\usepackage{textcomp}
\usepackage{amssymb}

\begin{document}

\title[Variability and stability in blazar jets]{Variability and stability in blazar jets on time scales of years: \\ Optical polarization monitoring of OJ287 in 2005--2009}

\author[C. Villforth et al.]{\large C.Villforth$^{1,2}$\thanks{E-mail:
carovi@utu.fi}, K.Nilsson$^{1}$, J.Heidt$^{3}$, L.O.Takalo$^{1}$, T.Pursimo $^{2}$, A.Berdyugin$^{1}$, E.Lindfors$^{1}$, M.Pasanen$^{1}$, M.Winiarski$^{4}$, M.Drozdz$^{4}$, \newauthor \large W.Ogloza$^{4}$, M.Kurpinska-Winiarska$^{5}$, M.Siwak$^{5,6}$, D.Koziel-Wierzbowska$^{5}$, C.Porowski$^{5}$, A.Kuzmicz$^{5}$, J.Krzesinski$^{4}$, T.Kundera$^{5}$, \newauthor \large J.-H.Wu$^{7}$, X.Zhou$^{7}$, Y.Efimov$^{8}$, K.Sadakane$^{9}$, M.Kamada$^{9}$, J.Ohlert$^{10}$, V.-P.Hentunen$^{11}$, M.Nissinen$^{11}$, M.Dietrich$^{12}$, R.J.Assef$^{12}$,\newauthor \large D.W.Atlee$^{12}$, J.Bird$^{12}$, D.L.DePoy$^{13}$, J.Eastman$^{12}$, M.S.Peeples$^{12}$, J.Prieto$^{12}$, L.Watson$^{12}$, J.C.Yee$^{12}$, A.Liakos$^{14}$, P.Niarchos$^{14}$, \newauthor \large K.Gazeas$^{14}$, S.Dogru$^{15}$, A.Donmez$^{15}$, D.Marchev$^{16}$, S.A.Coggins-Hill$^{17}$, A.Mattingly$^{18}$, W.C.Keel$^{19}$, S.Haque$^{20}$, \newauthor \large A.Aungwerojwit$^{21,22}$ and N.Bergvall$^{23}$\\
$^{1}$Tuorla Observatory, Department of Physics and Astronomy, University of Turku, V\"{a}is\"{a}l\"{a}ntie 20, FI-21500 Piikki\"{o}, Finland\\
$^{2}$Nordic Optical Telescope, Apartado 474, E-38700 S/C de la Palma, Spain\\
$^{3}$ZAH, Landessternwarte Heidelberg, K\"{o}nigstuhl, 69117 Heidelberg, Germany\\
$^{4}$Mt. Suhora Observatory, Pedagogical University, ul. Podchorazych 2, 20-084 Krakow, Poland\\
$^{5}$Astronomical Observatory, Jagiellonian University, ul. Orla 171, 30-244 Krakow, Poland\\
$^{6}$Department of Astronomy and Astrophysics, University of Toronto, 50 St. George St., Toronto, Ontario, M5S\~3H4, Canada\\
$^{7}$National Astronomical Observatories, Chinese Academy of Sciences, 20A Datun Road, Beijing 100012, China\\
$^{8}$Crimea Astrophysical Observatory, Yalta, 334242 Crimea, Ucraine\\
$^{9}$Astronomical Institute, Osaka-Kyoiku University, Asahigaoka, Kashiwara, Osaka 582-8582, Japan\\
$^{10}$Michael Adrian Observatorium, Astronomie Stiftung Trebur, Fichtenstrasse 7, 65468 Trebur, Germany\\
$^{11}$Taurus Hill Observatory, H\"{a}rk\"{a}m\"{a}entie 88, FI-79480, Kangaslampi Finland\\
$^{12}$Department of Astronomy, The Ohio State University, 4055 McPherson Lab, 140 W. 18th Ave, Columbus, OH 43210 U.S.A.\\
$^{13}$Department of Physics and Astronomy, Texas A\&M University, College Station, TX 77843, U.S.A.\\
$^{14}$Department of Astrophysics, Astronomy and Mechanics, Faculty of Physics, University of Athens, Panepistimiopolis, GR-15784 Zografos, Athens, Greece\\
$^{15}$Canakkale Onsekiz Mart University, Faculty of Physics, TR-17020 Canakkale, Turkey\\
$^{16}$Department of Physics, Shoumen University, 9700 Shoumen, Bulgaria\\
$^{17}$Am Weinberg 16, 63579 Freigericht-Horbach, Germany\\
$^{18}$Grove Creek Observatory, Trunkey Creek, NSW 2796, Australia\\
$^{19}$Department of Physics and Astronomy, University of Alabama, Tuscaloosa, AL 35487-0324, USA\\
$^{20}$Department of Physics, University of West Indies, St. Augustine, Trinidad\\
$^{21}$Department of Physics, Faculty of Science, Naresuan University, Phitsanulok, 65000, Thailand\\
$^{22}$Department of Physics, University of Warwick, Coventry CV4 7AL, UK\\
$^{23}$Department of Astronomy and Space Physics, Uppsala University, Box 515, 751 20 Uppsala, Sweden\\
}
\date{Accepted: 27 November 2009. Received: 27 October 2009}
\pagerange{\pageref{firstpage}--\pageref{lastpage}} \pubyear{2009}

\maketitle

\label{firstpage}

\begin{abstract}
\small OJ287 is a BL Lac object at redshift $z=0.306$ that has shown double-peaked bursts at regular intervals of $\sim 12$ yr during the last $\sim 40$ yr. We analyse optical photopolarimetric monitoring data from 2005--2009, during which the latest double-peaked outburst occurred. The aim of this study is twofold: firstly, we aim to analyse variability patterns and statistical properties of the optical polarization light-curve. We find a strong preferred position angle in optical polarization. The preferred position angle can be explained by separating the jet emission into two components: an optical polarization core and chaotic jet emission. The optical polarization core is stable on time scales of years and can be explained as emission from an underlying quiescent jet component. The chaotic jet emission sometimes exhibits a circular movement in the Stokes plane. We find six such events, all on the time-scales of 10--20 days. We interpret these events as a shock front moving forwards and backwards in the jet, swiping through a helical magnetic field. Secondly, we use our data to assess different binary black hole models proposed to explain the regularly appearing double-peaked bursts in OJ287. We compose a list of requirements a model has to fulfil to explain the mysterious behaviour observed in OJ287. The list includes not only characteristics of the light-curve but also other properties of OJ287, such as the black hole mass and restrictions on accretion flow properties. We rate all existing models using this list and conclude that none of the models is able to explain all observations. We discuss possible new explanations and propose a new approach to understanding OJ287. We suggest that both the double-peaked bursts and the evolution of the optical polarization position angle could be explained as a sign of resonant accretion of magnetic field lines, a 'magnetic breathing' of the disc.
\end{abstract}

\begin{keywords}
accretion, accretion discs -- magnetic fields -- polarization -- shock waves -- BL Lacartea objects : individual : OJ287 --  galaxies : jets
\end{keywords}

\section{INTRODUCTION}
\label{sec:intro}

Blazars are amongst the most violently variable sources in the Universe. According to the standard model \citep{urry_unified_1995}, blazars are AGN with a jet pointing almost directly towards the observer, the jet radiation is thus highly beamed and dominates the spectrum. Therefore, blazars are perfect laboratories to study variability and turbulence in AGN jets.

Blazars are divided into two subclasses: BL Lac objects and Flat Spectrum Radio Quasars (FSRQs). Both BL Lacs and FSRQs show a flat radio spectrum, high polarization and violent variability \citep{urry_unified_1995}. While BL Lacs show a featureless optical spectrum with extremely weak or absent broad and narrow emission lines, FSRQs show a normal spectrum of broad and narrow emission lines. In the standard interpretation, BL Lacs are thought to be the equivalents of Fanaroff-Riley I (FR I) radio galaxies at small viewing angle, while FSRQs are thought to be the equivalents of Fanaroff-Riley II (FR II) radio galaxies \citep{ghisellini_fermi_2009}. Lately, more and more evidence has been found to support the idea that FSRQs and BL Lac differ not only in the appearance of their jets on kpc scale but also in their accretion processes. It has been suggested that FSRQs accrete in a radiatively efficient, geometrically thin accretion disc, while BL Lacs accrete through radiatively inefficient accretion flows (RIAFs) \citep{baum_toward_1995, ghisellini_fermi_2009}.  

However, even FSRQs and BL Lacs might differ in their accretion process and appearance of the jets, the variability observed in both types of objects is similar \citep{ulrich_variability_1997}. Both classes show variability on time scales from hours to decades, with partially extreme amplitudes \citep{ulrich_variability_1997,valtaoja_radio_2000,villforth_intranight_2009}. In both classes, the radio structure from VLBI maps consists of a so-called radio-core that does not move and blobs that appear to be ejected from the core and move away from it at apparently superluminal speeds. These blobs have been explained as signs of shock-fronts moving along the jet. Shock fronts are also thought to cause powerful outbursts lasting several weeks that can be observed on all wavelengths \citep{marscher_models_1985}. Using multi-wavelength data, it is possible to model the physical conditions in these shock fronts \citep{marscher_models_1985,marscher_inner_2008,darcangelo_synchronous_2009}.

Shorter time-scales and the behaviour in the quiescent phases are however less well understood. Studying variability on shorter time-scales is a very challenging field. On these time-scales, multi-wavelength observations cannot be used due to the fact that the delay between the different wavelength and the variability itself appear on similar time-scales. Neither are these short time-scales and thus small angular movements accessible using VLBI maps. Even so, it is a well known fact that variability down to the time-scales of hours exists \citep{wagner_intraday_1995,ulrich_variability_1997,villforth_intranight_2009}. Using optical polarization data, it is possible to study the variability in blazar jets down to the smallest time-scales, and opposed to normal flux monitoring, polarization monitoring can reveal the evolution of the magnetic field.

Another open question, to which flux monitoring cannot answer, is what happens in the jet during quiescent phases. Using optical polarization monitoring, we can assess the properties of the magnetic field in the jet during quiescent phases and can thus answer the question if stable components in the jet emission exist.


For our study, we choose OJ287, which is one of the best studied blazars. It has been monitored since the last century accidentally and has been studied excessively since 1970s both using photometry and polarimetry. Therefore, OJ287 is a perfect object to study variability in blazar jets on a large range of time-scales, from weeks up to decades. OJ287 has a moderate redshift of $z=0.306$ \citep{sitko_continuum_1985}.

OJ287 has received a lot of attention as it has shown massive double-peaked outbursts approximately every 12 years during the last 40--100 years. Such long-lasting, regularly appearing events have not been observed in any other AGN so far. \citet{sillanpaa_oj_1988} first noted this exceptional behaviour and suggested that the regularly appearing outbursts might be caused by a close binary black hole system in which the secondary black hole induces tidal disturbances in the accretion disc of the primary black hole. However, due to the limited amount of data, \citet{sillanpaa_oj_1988} were not able to explain how exactly the flare happens.

\citet{lehto_oj_1996} further developed this model and suggested that each of the flares actually constitutes of two flares: the first flare is extremely short and is caused by the crossing of the secondary black hole through the disc, while the second flare is caused by enhanced accretion due to tidal disturbances in the accretion disc. The time lag between the two flares is extremely short ($\sim$ 1 week), so that they are actually observed as one big outburst. Two of those outbursts happen during each orbit of the secondary black hole. The orbit is extremely eccentric, so that we observe two outbursts separated by only about 1 year every 12 years. \citet{lehto_oj_1996} modelled the previous behaviour and predicted the next outburst to happen in Spring 2006. However, OJ287 did not fail to surprise observers, and the first outburst in the 2005--2007 season happened about half a year earlier than predicted by \citet{lehto_oj_1996}. After the early 2005 burst, this model was further modified to fit the newest data \citep{valtonen_predictingnext_2006,valtonen_new_2007,valtonen_tidally_2009}.

Several other authors have suggested models for OJ287 \citep{katz_precessing_1997,villata_beaming_1998,valtaoja_radio_2000}, all of them based on the assumption that OJ287 hosts a close binary black hole. \citet{katz_precessing_1997} suggested that a binary black hole induces precession in the accretion disc of OJ287, the jet follows the precessing motion of the discs and sweeps through the line of sight regularly, causing major flares due to enhanced beaming. \citet{villata_beaming_1998} suggested that OJ287 hosts a binary black hole, in which both black holes produce a jet, the jets sweep through the line of sight on regular intervals causing double-peaked bursts.

\citet{valtaoja_radio_2000} studied radio monitoring data of OJ287 and noted differences between the first and second burst and therefore suggested that the first burst is caused by a disc-crossing while the second burst is related to enhanced accretion causing a shock front in the jet. Using the radio data, \citet{valtaoja_radio_2000} further argued that the missing of radio counterparts in some bursts speaks against the beaming models \citep{katz_precessing_1997,villata_beaming_1998} as for beaming events one would expect all wavelengths to be enhanced in a similar manner. The observed burst however do not show such a behaviour.

Both the Lehto \& Valtonen, its modified successor \citep{valtonen_predictingnext_2006,valtonen_new_2007} and the Valtaoja model make clear predictions about the appearance of the bursts in polarization and the exact timing of the bursts. Thus, our data sets can be used to test the different models.

In this paper we analyse optical photopolarimetric monitoring of OJ287 during 2005--2009. We analyse the variability in optical polarization and compare it to current blazar jet models. We compare the data to different models proposed to explain the regularly appearing outbursts in OJ287. In Section 2 we present the observations and data reduction. Results are presented in Section 3. We discuss our results, both concerning the general variability pattern observed and the different models for OJ287, in Section 4. Conclusions are presented in Section 5. The cosmology used is $H_{0}=70\textrm{km s}^{-1}\textrm{Mpc}^{-1}, \Omega_{\Lambda}=0.7, \Omega_{m}=0.3$.

\section{OBSERVATIONS AND DATA REDUCTION}
\label{sec:observations}

We have obtained 400 polarimetric and 2238 photometric observations of OJ287 in 2002--2009. The participating observatories and distribution of data points among them is shown in Table \ref{observatories}.

Of the 400 polarimetric measurements 110 were taken in service mode using the focal reducer CAFOS at the Calar Alto (CA) 2.2 m telescope.  A rotatable $\lambda$/2 plate + Wollaston prism were employed and four exposures through the R-band were taken with the $\lambda$/2 plate rotated by $22.5\degr$ between the exposures. Exposure times were typically 60 s.  The field of view was large enough (7$\times$7 arcsec) to include a number of comparison stars from \citet{gonzlez-prez_optical_2001} enabling simultaneous photometric measurements of OJ287. Data were bias-subtracted and flatfielded using full polarimetric flatfields.

Altogether 256 polarimetric measurements were taken with the remotely controlled 60 cm KVA telescope on La Palma. The integration times were similar to the Calar Alto observations, but in order to improve the signal-to-noise, four to 12 sequences of four images were taken and the individual polarization measurements were averaged. Furthermore, the images were made in \textquotedblleft white light\textquotedblright, i.e. without a filter. A calcite plate was used instead of a Wollaston prism to separate the ordinary and extraordinary beams. Polarimetric data were dark-subtracted. Nearly simultaneous (time difference $<$ 30 min) R-band photometry was obtained from CCD images made with a 35 cm telescope attached to the 60 cm telescope.

Finally, 34 polarimetric measurements were made with the Nordic Optical Telescope (NOT) using an identical setup to the KVA, except that R-band filter was used. Data were bias-subtracted and flatfielded using full polarimetric flatfields. Simultaneous \textit{R}-band photometry was derived from star 13 in \citet{gonzlez-prez_optical_2001}.

The normalized Stokes parameters $P_Q$ and $P_U$ and the degree of polarization $P$ and position angle $PA$ were computed from the intensity ratios of the ordinary and extraordinary beams using standard formulae (see e.g. \citealt{landi_deglinnocenti_polarimetric_2007}) and semi-automatic software specifically developed for polarization monitoring purposes. During some of the nights, polarized standard stars from \citet{turnshek_atlas_1990} were observed to determine the zero point of the position angle. The instrumental polarization was found to be negligible for all three telescopes. Foreground (Galactic) polarization should also be low ($<$ 0.5 per cent) since the reddening value is only ${\rm E_{\rm B-V}}$ = 0.028. The degree of polarization was corrected for bias using the maximum likelihood estimator from \citet{simmons_point_1985}. Since the KVA observations were made in \textquotedblleft white light\textquotedblright the effective
wavelength is mainly determined by the sensitivity of the detector. The sensitivity of the Marconi 47-10 detector used at the KVA peaks at $\lambda \sim$ 500 nm, somewhat shorter than the peak of the \textit{R}-band filter used at CA and NOT ($\sim 640$ nm).  Since frequency-dependent polarization is commonly observed in OJ287 (see e.g. \citealt{holmes_polarization_1984}), a small offset between the Stokes parameters observed by CA and KVA is expected. However, the difference in effective wavelength is relatively small and any offsets are likely to be small too. We have examined the 24 cases where nearly simultaneous (time difference less than two hours) data have been obtained with both CA and KVA telescopes. We computed the mean of the Stokes parameters $P_Q$ and $P_U$ for both CA and KVA data sets and found a difference between the means to be $0.47\pm0.17$ and $0.76\pm0.13$ for $P_Q$ and $P_U$, respectively. These are very small offsets compared to the total range of $P_Q$ and $P_U$ and of the order of the error bars of a single point.  Thus differences in effective wavelength between the CA and KVA data do not cause large enough offsets to affect our conclusions.

In addition to the polarimetric points, 2238 photometric points were obtained in \textit{BVRI} bands in 18 observatories over the world (see Table \ref{observatories}). CCD images were obtained of the OJ287 field and the frames were reduced in the usual way by first subtracting the bias and dark frames and then dividing by a flat-field frame. The photometry was performed in differential mode using star nine in \citet{gonzlez-prez_optical_2001} as the comparison star. Inter-observer offsets were checked using star 13 as a control and found to be generally small ($<$0.02 mag). The offsets derived from star 13 were applied to the magnitudes of OJ287 to bring all measurements to the same scale. Finally, one-hour averages were computed from the data.

All data are available at Vizier.

\begin{table*}
\begin{minipage}{180mm}
\caption{\label{observatories}The participating telescopes in the OJ287 campaign.  The second column gives the observatory abbreviation used in the data tables and the third column the mirror diameter. The last columns give the number of photometric and polarimetric data points.}
\begin{tabular}{lrrrrrrrr}
\hline
Observatory & abbr. & D & N$_U$ & N$_B$ & N$_V$ & N$_R$ & N$_I$ & N$_{\rm Pol}$\\
& & [cm] & &\\
\hline
Athens, Greece           & ATH &  40 & & &     &  79  &     &\\
Calar Alto, Spain        & CA  & 220 & & &     & 118  &     & 110\\
Canakkale, Turkey        & CO  &  40 & & &   3 &  12  &   5 &\\
Grove Creek, Australia   & GC  &  30 & & &     &   8  &     &\\
Heidelberg, Germany      & HEI &  70 & &  &     &   1  &     &\\
Krakow, Poland           & KR  &  50 & & 14 &  47 &  57  &  25 &\\
MDM, USA                 & MDM & 240 & & &     &  34  &     &\\
KVA, La Palma            & KVA &  35 & &  &     & 406  &     & 256\\
NOT, La Palma            & NOT & 256 &  & &     &  45  &     &  34\\
Osaka, Japan             & OSA &  51 &  &   &  63 &  89  &  71 &\\
SATU\footnote{St. Augustine-Tuorla}, Trinidad       & SAT &  & &     &     &   1  &     &\\
Mt. Suhora, Poland       & SUH &  60 & 5 & 101 & 178 & 208  & 143 &\\
Taurus Hill, Finland     & TH  &  30 &  &   &     &  49  &     &\\
Trebur, Germany          & TRE & 120 &   &  &     &  74  &     &\\
Tuorla, Finland          & TUO & 103 &   &  &     & 131  &     &\\
UA\footnote{University of Alabama}, USA              & UA  &  40 &  &   &     &   2  &     &\\
Xinglong, China          & XIO &  90 & &  &     & 159  &     &\\
Liverpool Telescope, La Palma  & LIV  & 200 & & 24 & 24 &38\footnote{SDSS 'r} &24\footnote{SDSS 'i} &\\
Total                    &     &     & 5 & 139 & 315 & 1511 & 268 & 400\\
\hline
\end{tabular}
\end{minipage}
\end{table*}

\section{RESULTS}
\label{sec:results}

\subsection{General appearance of the light-curve: the major outbursts}
\label{sec:double-bursts}

\begin{figure*}
\begin{minipage}{18cm}
\includegraphics[width=18cm]{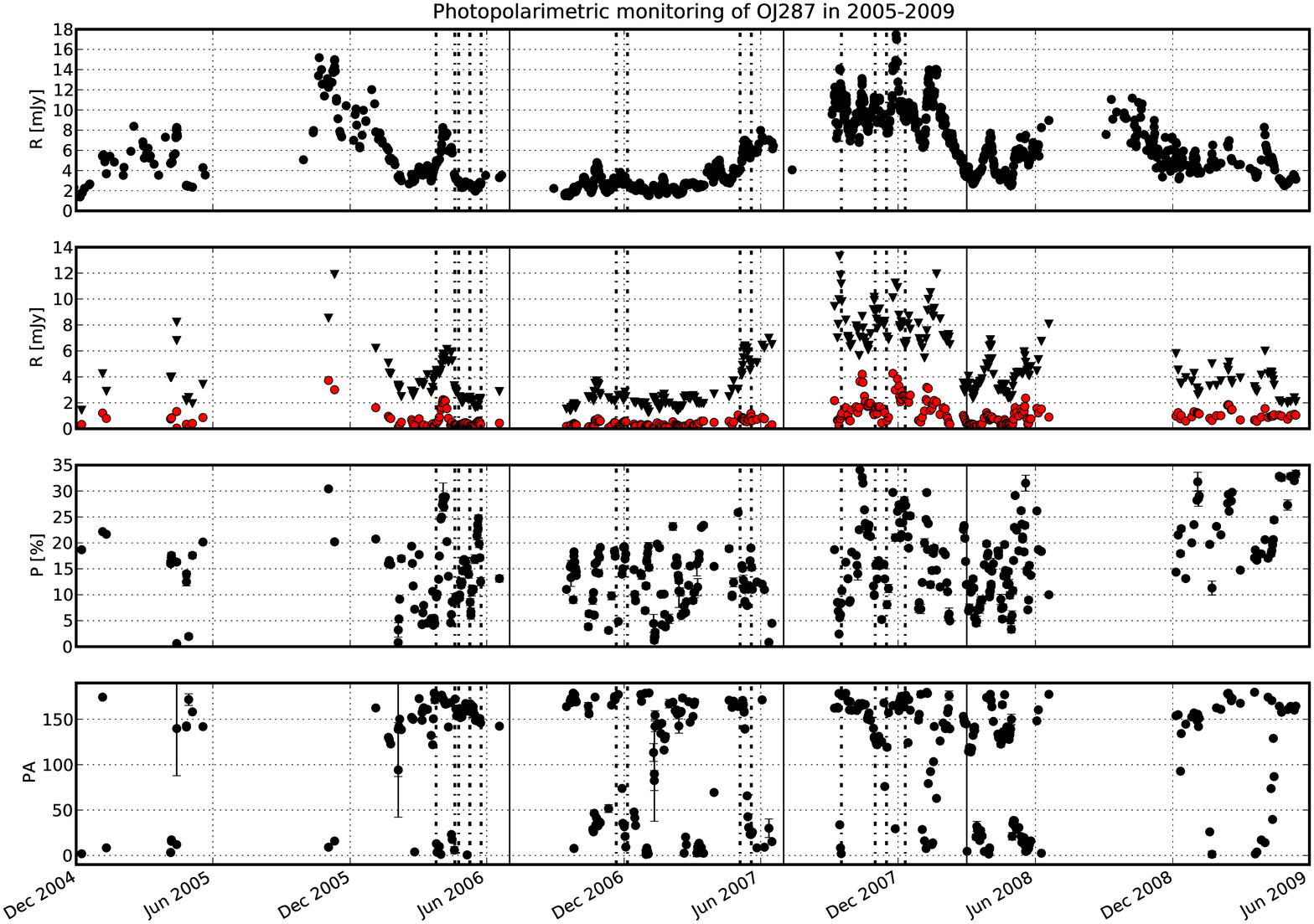}
\caption{Full photopolarimetric light-curve for OJ287 in 2005--2009. Panels show following values, from top to bottom: total flux; polarized (red circles) and unpolarized (black triangles) flux; degree of polarization; position angle. Solid vertical lines show data bins for OPC variability used in Section \ref{sec:lightcurvestats}. Vertical dash-dotted lines show beginning and end of bubbles and swings as discussed in Section \ref{sec:bubbles}.}
\label{alldata}
\end{minipage}
\end{figure*}

We start by analysing and discussing the appearance of the light-curve in general and describe the two major outbursts observed during 2005--2009. The full light-curve in photometry and polarimetry is presented in Fig. \ref{alldata}. The gaps in the light-curves appearing in June--July yearly are due to the fact that OJ287 is too close to the sun during these months. We will refer to these gaps as 'summer gaps'. 

\begin{figure*}
\begin{minipage}{18cm}
\includegraphics[width=18cm]{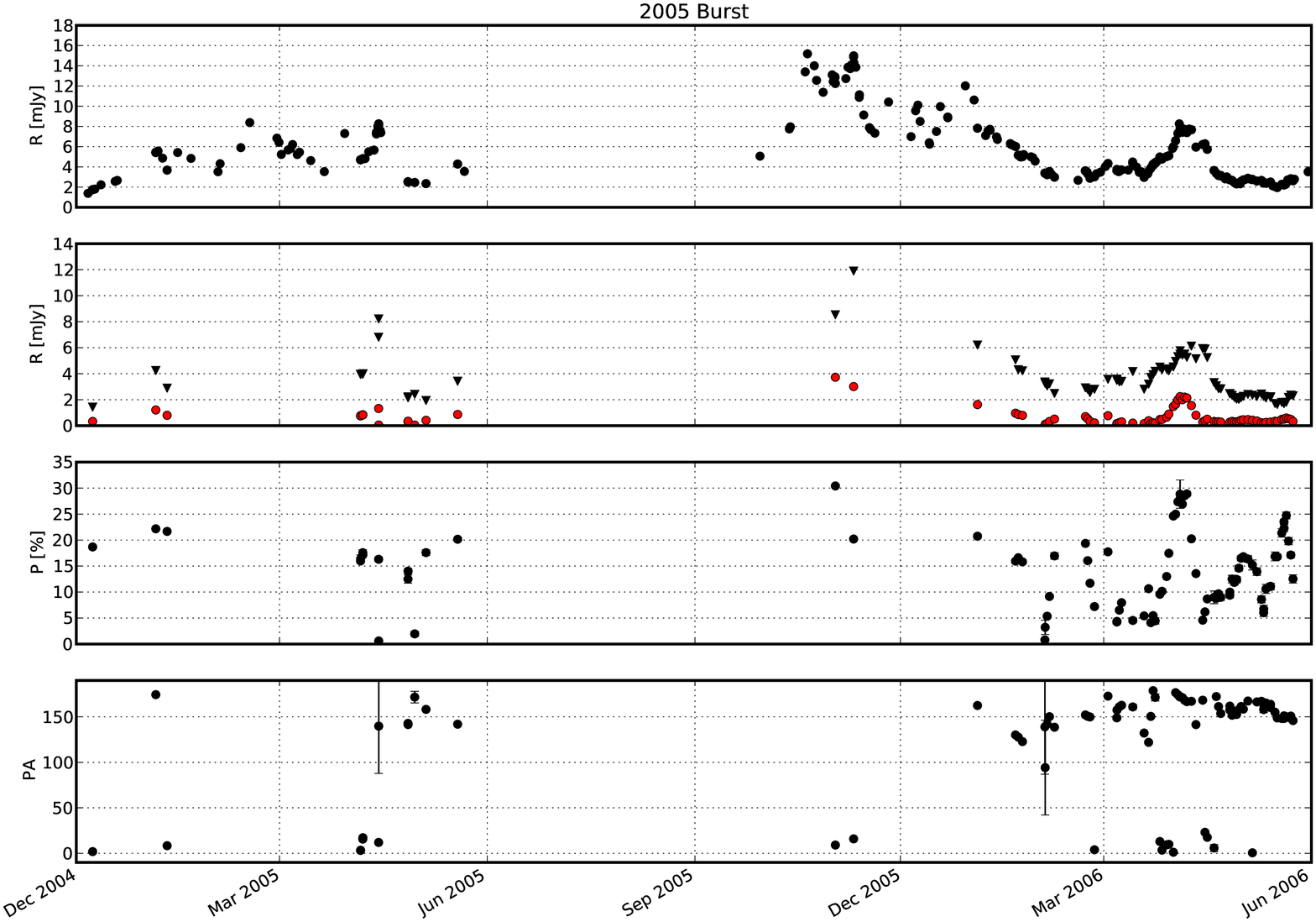}
\caption{Detailed view of the 2005 burst. Panels show following values, from top to bottom: total flux; polarized (red circles) and unpolarized (black triangles) flux; degree of polarization; position angle.}
\label{2005burst}
\end{minipage}
\end{figure*}

\begin{figure}
\includegraphics[width=8cm]{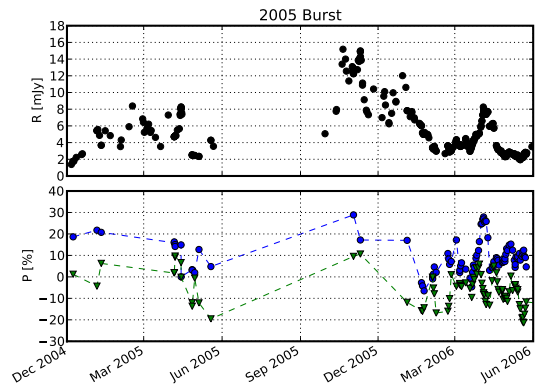}
\caption{Normalised Stokes parameters during the 2005 burst. Upper panel: flux, lower panel: $P_{Q}$ (blue circles) and $P_{U}$ (green triangles).}
\label{2005stokes}
\end{figure}

\begin{figure*}
\begin{minipage}{18cm}
\includegraphics[width=18cm]{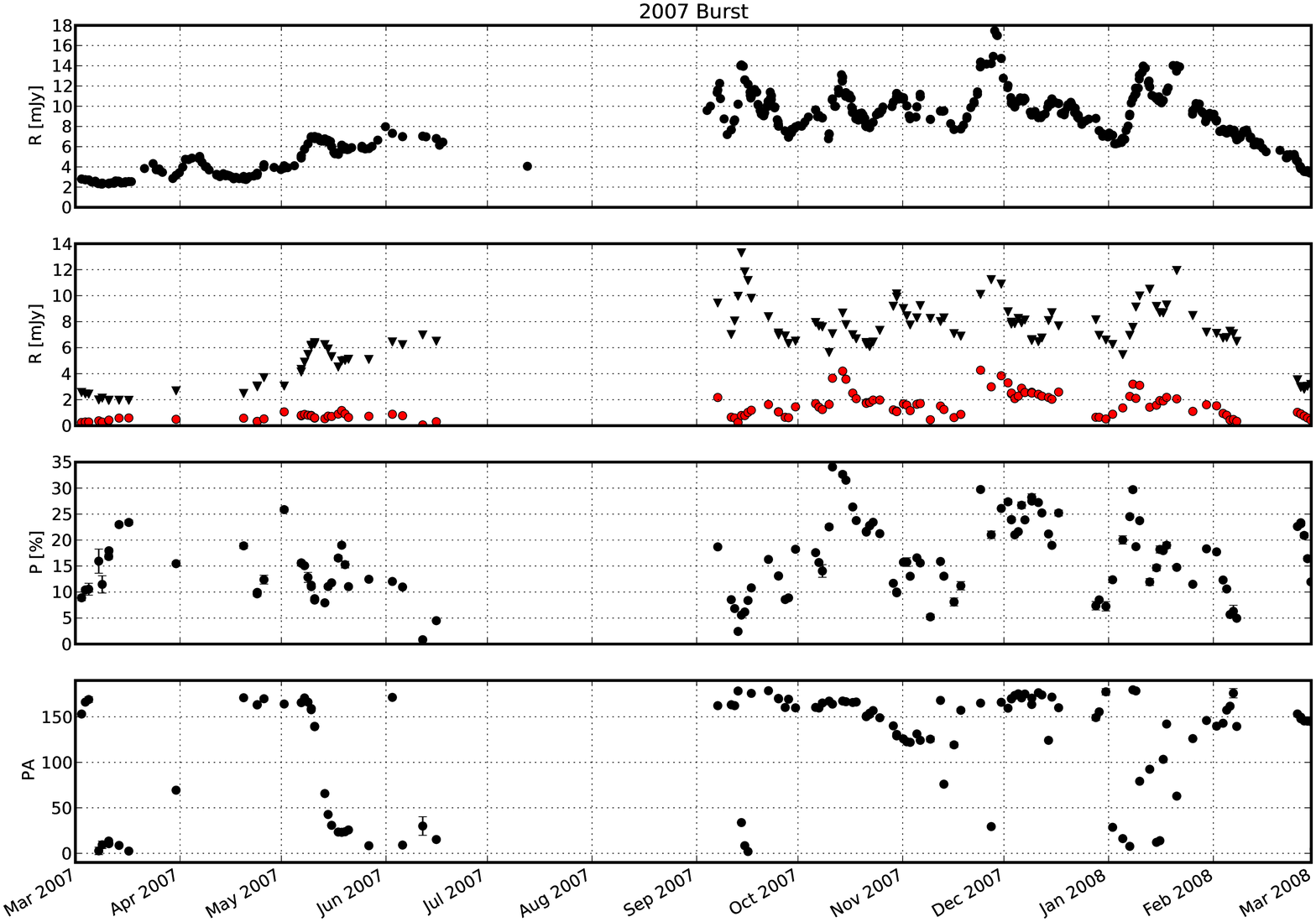}
\caption{Detailed view of the 2007 burst. Panels show following values, from top to bottom: total flux; polarized (red circles) and unpolarized (black triangles) flux; degree of polarization; position angle.}
\label{2007burst}
\end{minipage}
\end{figure*}

\begin{figure}
\includegraphics[width=8cm]{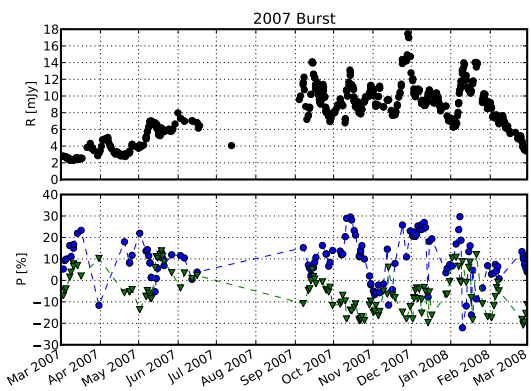}
\caption{Normalised Stokes parameters during the 2007 burst. Upper panel: flux, lower panel: $P_{Q}$ (blue circles) and $P_{U}$ (green triangles).}
\label{2007stokes}
\end{figure}

During our monitoring campaign, two major outbursts were observed, one in late 2005 and the other one in late 2007. Detailed plots of the two major outbursts are presented in Figures \ref{2005burst} (full data),  \ref{2005stokes} (Stokes parameters)  for the 2005 burst and \ref{2007burst} (full data), \ref{2007stokes} (Stokes parameters) for the 2007 burst.

The first major outburst is observed in late 2005. We do not see any exceptional behaviour before the 'summer gap' in 2005. The flux is relatively low and the polarization is moderately high (P$\sim$ 20 per cent). After the 'summer gap', in late 2005, we see a rather smooth increase in flux, the polarization is similarly high as before the burst. The main flare shows two major bursts of similar strength, separated by about a month. The object stays in burst till early 2006, the variability after the two peaks is erratic. A smooth decline starts in spring 2006. In the end phase of the decline, when the flux almost reaches a normal level again, a short, highly polarized outburst occurs. After this short flare, the source returns to normal flux levels when the 2006 'summer gap' sets in.

As for the exact timing of the burst, we see two major flares of similar strength in this burst. It is unclear which of the two is the main flare. The first of the two flares is observed on 20 October 2005 and the second flare is observed on 9 November 2005. The error in these values are in the order of days due to sampling limitations.

This burst was predicted to happen in 2006 by both \citet{lehto_oj_1996} (2006 May 12) and \citet{valtaoja_radio_2000} (2006 September 25). Thus, we can clearly say that our finding disagrees strongly with both existing predictions. The Lehto \& Valtonen model has an error of about six months, while the Valtaoja model has an error of almost a year. The modified Lehto \& Valtonen model \citep{valtonen_predictingnext_2006,valtonen_new_2007} by definition fits the burst in 2005 as it is based on the timing from that burst.

As for the appearance in polarization, only very few data points are available in polarization during the 2005 flare. No one expected a major burst at this point and thus few telescopes were monitoring OJ287 in polarization. From the few data points we have, it seems as if the whole burst was rather strongly polarized. Only two data points were observed during the flares, both showing rather high degrees of polarization. The position angle in those two data points is around zero, which is rather close to the value of $\sim$ 170\degr, which the position angle is observed to fluctuate around most of the time.

After the 2006 'summer gap', a moderately polarized, steep and short outburst is observed. Whereupon the light-curve is without prominent features, only showing smaller, rather erratic bursts. Already in early June 2007, the flux starts to increase rather steeply, announcing the beginning of the second major outburst. The polarization during this rise is extremely low, the increase in flux is not smooth as for the 2005 burst, but overlaid with several smaller flares. After the 'summer gap', in September 2007, the outburst reaches its peak, the polarization is high and the variability in flux is erratic. The plateau of the burst lasts till January 2008. During this time, the flux stays on a constantly high level, superimposed with countless fast erratic flares. A smooth, rather steep decline begins in January 2008. The decline is interrupted by a highly polarized, sudden and steep flare, similar to the one observed after the 2005 burst. However, after the second burst, OJ287 does not return to its normal state, another highly polarized, broad and strong burst begins which lasts until the end of our monitoring campaign in summer 2009. The 'third major burst' is slightly less luminous ($\sim$12 mJy) than the first ($\sim$14 mJy) and second bursts ($\sim$16 mJy). The burst is extremely chaotic, similar to the second burst in its plateau.

While the 2005 burst was a test for the 'old' \citet{lehto_oj_1996} model and the \citet{valtaoja_radio_2000} models, the 2007 burst will only be used to test the 'new' Lehto \& Valtonen model \citep{valtonen_predictingnext_2006,valtonen_new_2007} as both other models did not predict the first burst correctly. The 'summer gap' poses a serious problem for the exact timing of the 2007 burst. We see a sharp rise of flux right before the 'summer gap'. When OJ287 is observable again, the outburst is already in its plateau. \citet{valtonen_massive_2008} identified the second flare after the summer gap as the major burst. This burst is not the highest peak in the plateau, but it shows extremely low levels of polarization, as predicted by \citet{valtonen_new_2007}. It is not clear if another strong peak occurred shortly before OJ287 returned from the 'summer gap'. A single, rather low photometric data-point was observed during the summer gap on 12 July 2007. Thus we can assume that the highest peak did not occur shortly after that date. We assume that several flares must have occurred before OJ287 returned to high luminosities from such a dip. As the time scale of the erratic flares is usually about a week, we conclude that the major outburst in 2007 has occurred no earlier than mid to late August 2007. The first peak after the summer gap was observed on 2007 September 12. However, several flares of similar strength are observed after that, the highest of those as late as December 2007. Thus, we cannot clearly say which of those flares is the main flare, it might have happened at any time between August and December 2007. The second flare after the summer gap is well consistent with predictions from the modified Lehto \& Valtonen model \citep{valtonen_predictingnext_2006,valtonen_new_2007}, who predicted the burst to happen in 2007 September 13. It is also practically unpolarized, as predicted by \citet{valtonen_new_2007}.

As for the appearance of the burst in polarization, while the degree of polarization starts off rather low during the beginning of the burst, it rises to extremely high values in the first post-maximum peak. The position angle stays relatively stable for the first two peaks after the summer gap while a swinging behaviour is seen in the third peak after the summer gap after which the position angle falls back to the preferred value of $\sim$170\degr. In Fig. \ref{2007stokes}, we plot the normalised Stokes parameters for the same time span, the two normalised Stokes parameters show similar evolution. Just as the photometric light-curve, the polarimetric light-curves shows rather erratic behaviour.

\subsection{Statistical properties of the polarimetric dataset}
\label{sec:lightcurvestats}

In this section, we will analyse the data using statistical methods. We will not consider single events in the light-curve, but the properties of the entire data set. Whenever we calculate the mean or standard deviation for the position angle, we use the circular mean and circular standard deviation \citep{mardia_directional_1975}. Throughout this paragraph, we use the Fisher kurtosis, normalized to zero. For all statistical tests we use the statistical packages \textit{stats} and \textit{morestats} from SciPy\footnote{http://www.scipy.org/}.

\begin{figure}
\includegraphics[width=8cm]{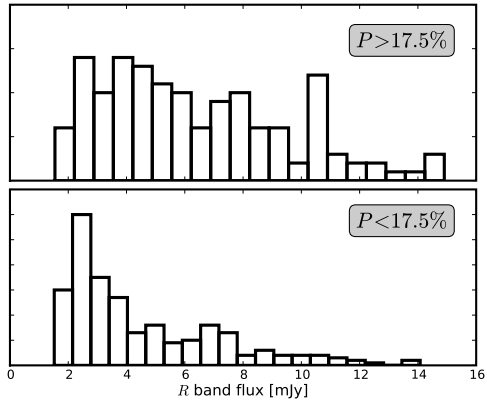}
\caption{Flux histogram for high ($P > $ 17.5 per cent, upper panel) and low polarization ($P < $ 17.5 per cent, lower panel).}
\label{Pvsflux_hist}
\end{figure}

First we would like to assess if the degree of polarization correlates with the \textit{R} band flux. While there is no strong correlation between the degree and the polarized flux, higher degrees of polarization tend to avoid very low fluxes, while low degrees of polarization are very common at low fluxes of $\sim$ 2 mJy. To assess this difference, we divide the sample into a high and low polarization sample. The histograms of the high and low polarization sample are shown in Fig. \ref{Pvsflux_hist}. We set the threshold to $P = 17.5$ per cent (the highest degree of polarization observed is $P = 34.06$ per cent, our threshold is about half of that value). We perform a Kolmogorov--Smirnov test, the probability that the two samples are drawn from the same parent populations but the differences are random is only $4.4\times 10^{-8}$. The difference between the low and high polarization sample is that the high polarization sample is missing the low flux peak at $\sim$ 2 mJy, which can be interpreted as a lower limit for the unpolarized flux. Our findings agree well with those of other authors \citep{jannuzi_optical_1994}.

\begin{figure}
\includegraphics[width=8cm]{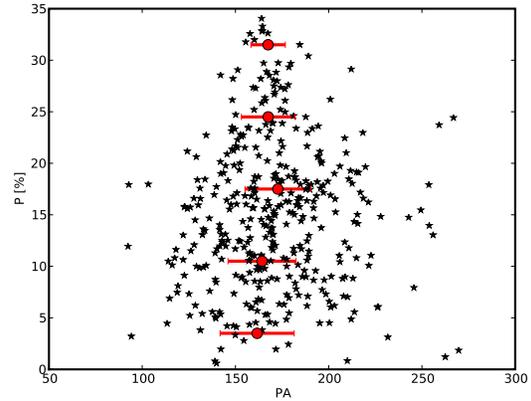}
\caption{Degree of polarization versus position angle for data 2005-2009. Filled circles with bars represent the circular mean and standard deviation of binned data.}
\label{PvsPA}
\end{figure}

As can be seen in Fig. \ref{alldata} the position angle typically lies around a value of $\sim$170\degr. The circular mean and standard deviation of the position angle is 167.7$\pm$17.5\degr. Thus, the optical polarization is oriented perpendicular to the radio jet \citep{darcangelo_synchronous_2009}. To investigate this further, we plot the degree of polarization over PA. As the PA has a range of 0--180 with 0 and 180 being 'identical', we calibrated the PA for this plot such that if $PA < 90$ we add 180 degrees to accomplish a PA range 90--270. The plot is shown in Fig. \ref{PvsPA}, a clear preferred position angle is visible. There is a seeming stronger alignment with higher degrees of polarization. Such a behaviour has been observed before by \citet{hagen-thorn_oj_1980} who studied OJ287 during 1971-1976. However, while in our case the preferred PA is $\sim$170\degr, \citet{hagen-thorn_oj_1980} find a preferred PA of 90\degr.

Using statistical tests, we will assess if the stronger alignment for higher degrees of polarization can be quantified. We divide the sample into several sub-samples and plot the mean and standard deviation. No significant decrease in the scatter is noted (Fig. \ref{PvsPA}). If we divide the data into two equally sized samples, one for high degrees of polarization and one for low degrees of polarization, the low polarization sample has a circular mean and standard deviation of 164.9$\pm$18.9\degr, while the high polarization sample has a circular mean and standard deviation of 170.0$\pm$15.8\degr. The values are consistent within the standard deviations of both averages, no significant difference between the two standard deviations is observed. However, if we perform Kolmogorov--Smirnov, the probability that the two samples are drawn from the same parent populations but the differences are random is 2.7 per cent. We try to investigate the origin of the seeming difference using the skew and kurtosis of the two samples. While the skew is similar for both samples, the kurtosis is significantly larger for the high polarization sample (1.6) than for the low polarization sample (0.4). Thus, while both distributions show a sharper peak and broader wings than a normal distribution, the high polarization sample is even more sharply peaked than the low polarization sample. This can be interpreted as a stronger alignment with higher degrees of polarization.

\begin{figure}
\includegraphics[width=8cm]{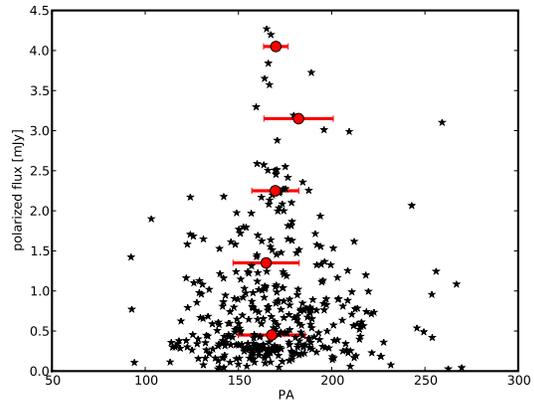}
\caption{Polarized flux versus position angle for data 2005--2009. Filled circles with bars represent the circular mean and standard deviation of binned data.}
\label{polfluxvsPA}
\end{figure}

Due to the fact that the degree of polarization is a parameter heavily contaminated by unpolarized flux, we perform the same analysis for the polarized flux (Fig. \ref{polfluxvsPA}). Just as for the degree of polarization, the alignment seems to get stronger, however, the scatter does not decrease significantly. The low polarized flux sample shows a circular mean and standard deviation of 165.2$\pm$18.0\degr while the high polarized flux sample shows a circular mean and standard deviation of 169.9$\pm$16.8\degr. But just as for the degree of polarization, if we perform a Kolmogorov--Smirnov test, the probability that the two samples are drawn from the same parent populations but the differences are random is 6 per cent. As for the degree of polarization, we tested for differences in the distributions. Again, the high polarization sample has higher kurtosis (1.3) than the low polarization sample (0.6).

Thus, for both the degree of polarization and the polarized flux there is a slight trend for the distributions to get narrower at higher polarization. This can be interpreted as stronger alignment at higher degrees of polarization.

\begin{figure}
\includegraphics[width=8cm]{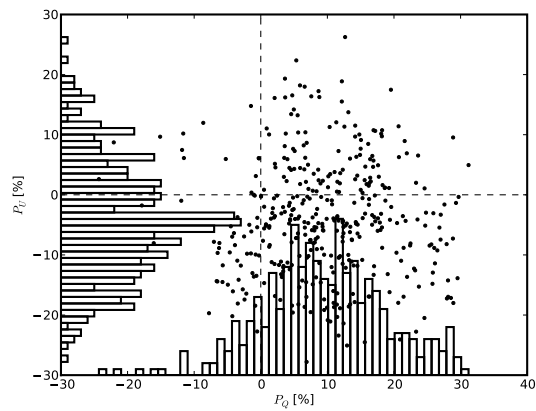}
\caption{Stokes plane plot for polarimetric data 2005--2009. Histograms of $P_{Q/U}$ are projected to the corresponding axes. Dashed lines indicate $P_{Q/U}=0$.}
\label{stokes}
\end{figure}

\begin{figure}
\includegraphics[width=8cm]{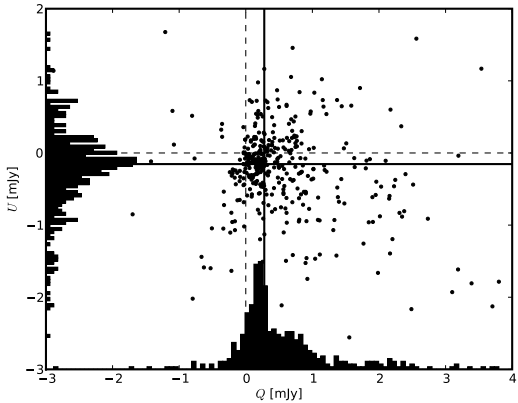}
\caption{Stokes plane plot in polarized flux $Q/U$ for polarimetric data 2005--2009. Histograms of $Q/U$ are projected to the corresponding axes. Dashed lines indicate $Q/U=0$, solid lines estimates for best values for $Q/U$, for details see text.}
\label{polfluxstokes}
\end{figure}

To investigate the reason behind the alignment, we present a Stokes plane plot for both $P_{Q/U}$ (Fig \ref{stokes}) and $Q/U$ (Fig. \ref{polfluxstokes}). To assess the distributions of $P_{Q/U}$ and $Q/U$, we project histograms of both values to the axes of the plots. Both $P_{Q/U}$ and $Q/U$ show preferred values, however, while the distributions in the degree of polarization are extremely broad, the distributions in the absolute Stokes parameters are much tighter. Therefore we will concentrate on the polarized flux from now on.

As we can already see in Fig. \ref{polfluxstokes}, the distribution of both $Q$ and $U$ are rather strongly skewed. $Q$ has a skew of 1.15, the peak lies at 0.28 mJy, thus $Q$ has a stronger tail on the side away from the zero point. $U$ has a skew of -0.22, the peak lies at -0.15 mJy, thus just as $Q$, $U$ has a stronger tail away from the zero point. To describe the shape of the distribution, we calculate the Fisher kurtosis. $Q$ has a kurtosis of 3.7, $U$ has a kurtosis of 1.6. Thus both distributions have extremely wide wings and sharp peaks, $Q$ is even more strongly peaked than $U$.
 
We interpret this finding such that there is an underlying, stable source of polarized emission, causing the peak, the optical polarization core (OPC). Using this interpretation, we separate the emission into two components:

\begin{itemize}
\item the optical polarization core (OPC)
\item turbulent, chaotic jet emission
\end{itemize}

Due to the strong skew in the distribution, arithmetic mean or median are not suitable to determine the exact value of the OPC. We therefore determine the values from sigma-clipped data ($\sigma$=2, 5 iterations). These estimates are plotted in Fig. \ref{polfluxstokes} as solid lines, we see that the values describe the peak of the distribution very well. The parameters of the OPC are: $Q$ = 0.28 mJy, $U$ = -0.15 mJy, total polarized flux = 0.32 mJy. For comparison, the flux distribution peaks at around 2.5 mJy. Thus the polarized OPC emission represents about 10 per cent of the quiescent total flux emission. Assuming that the OPC emission is maximally polarized ($P = 70$ per cent), the total flux of the OPC is $F_{OPC,total} = 0.46$ mJy. Thus for a common flux of 2.5 mJy, the OPC emission contributes about 20 per cent to the total flux.

Using the OPC values derived above, we calculate a core-subtracted polarized flux by subtracting the OPC vectorially from every data point. We then use the core-subtracted $Q/U$ to calculate a core-subtracted polarized flux and position angle.

\begin{figure}
\includegraphics[width=8cm]{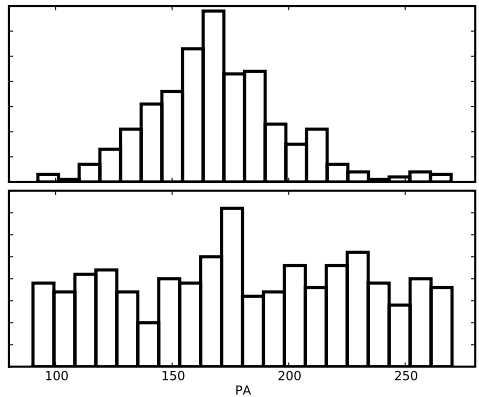}
\caption{Distributions of position angle $PA$ (upper panel) and core-subtracted position angle $PA$ (lower panel).}
\label{centeredPA}
\end{figure}

We use the core-subtracted position angle to assess if the alignment of the OPC persists in the turbulent jet emission. Fig. \ref{centeredPA} shows the comparison between the normal and core subtracted PA. The distribution of the core-subtracted PA is almost flat, we do not see a clear preferred position angle as in case of the raw data. The core subtracted position angle has a circular mean and standard deviation of 9.4 $\pm$ 27.8\degr, compared to 167.7$\pm$17.5\degr for the raw data. This is only a very weak alignment (no alignment would correspond to a standard deviation of $\sim$60\degr). The weak residual preferred position angle agrees with the original preferred PA within the errorbars.



\begin{figure}
\includegraphics[width=9cm]{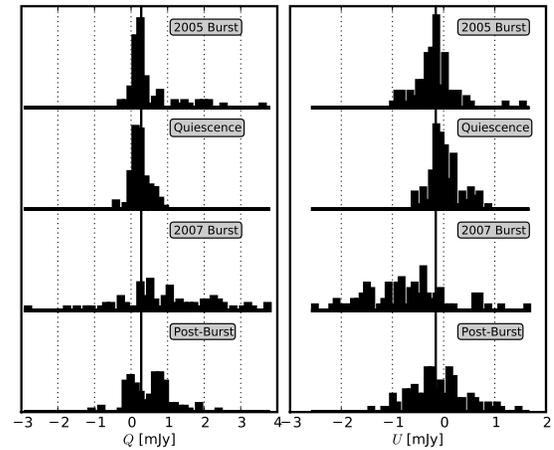}
\caption{Evolution of Stokes parameters during 2005--2009. Left panel: $Q$, right panel: $U$. All histograms are normed, same bins have been used for all histograms of a given Stokes parameter. Time spans for the different data bins are as follows: '2005 Burst' start of monitoring till 1 July 2006; 'Quiescence': 1 July 2006 -- 1 July 2007; '2007 Burst': 1 July 2007 -- 1 March 2008; 'Post-Burst': 1 March 2008 -- end of monitoring campaign. Data bins were chosen to reflect obvious periods in the light curve and so that different bins contain similar amounts of data points. Vertical black lines denote position of the OPC.} 
\label{OPC_evo_short}
\end{figure}

\begin{figure}
\includegraphics[width=8cm]{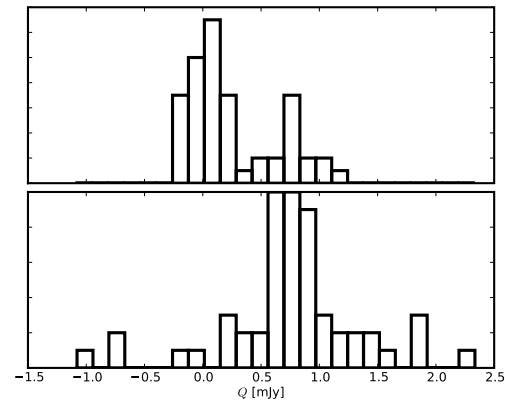}
\caption{Evolution of the OPC in Stokes $Q$ during the 'Post-Burst' phase. Upper panel shows first half of the data point, lower half shows second half of the data points. Histograms are normed, same bins are used for both histograms.}
\label{OPCjumps}
\end{figure}

\begin{figure}
\includegraphics[width=8cm]{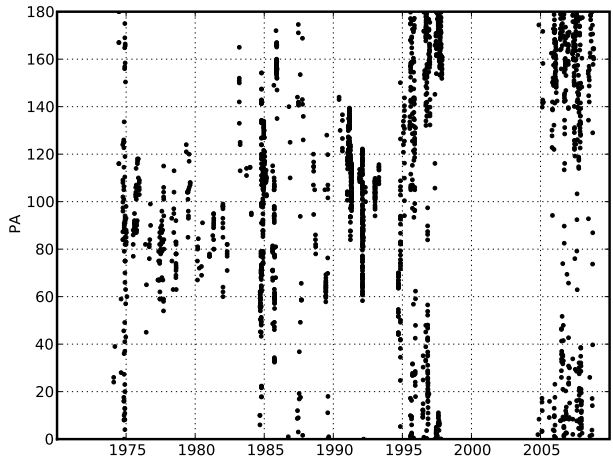}
\caption{Evolution of the optical position angle $PA$ of OJ287 from $\sim$ 1970 till today.}
\label{historicPA}
\end{figure}

\begin{figure}
\includegraphics[width=9cm]{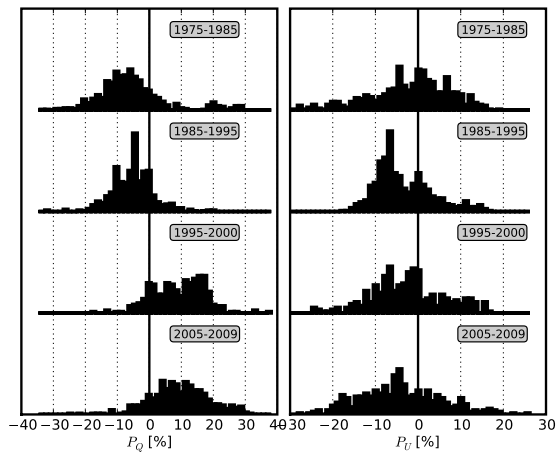}
\caption{Evolution of normalised Stokes parameters during 1975--2009. Left panel: $P_{Q}$, right panel: $P_{U}$. All histograms are normed, same bins have been used for all histograms of a given Stokes parameter. Solid vertical lines denote $P_{Q/U}$ = 0 per cent.} 
\label{OPC_evo_long}
\end{figure}

Next, we will assess if the OPC shows an evolution during our monitoring campaign. Therefore, we divide our sample into four data bins, the used bins are shown as solid vertical lines in Fig. \ref{alldata}. We choose the data bins to reflect major events in the optical light-curve. The four bins are:
\begin{itemize}
\item '2005 Burst' (till 1 July 2006): pre-burst and peak and decline of 2005 burst
\item 'Quiescence' (1 July 2006 -- 1 July 2007): phase between major bursts
\item '2007 Burst' (1 July 2007 -- 1 March 2008): plateau and decline of 2007 burst
\item 'Post-Burst' (from 1 March 2008): post-burst
\end{itemize}
All four bins contain approximately 100 data points. Normed histograms of the Stokes parameters for the four bins are presented in Fig. \ref{OPC_evo_short}


During the '2005 Burst', both Stokes parameters show strongly peaked distributions. The alignment becomes even strong during the 'Quiescence', when there is no power in the wings of the distributions. The OPC is constant during the '2005 Burst' and the 'Quiescence'. The picture changes dramatically with the '2007 burst', the alignment almost disappears, the distribution is extremely broad. It is very hard to estimate if the OPC is still at a similar value as before due to the large scatter. There is a tendency for the Stokes parameters to have the same sign, but be bigger than the OPC, this shows in both $Q$ and $U$. This might be the alignment that persisted during OPC-subtraction in Fig. \ref{centeredPA}.

The most spectacular bin however is the 'Post-Burst' bin. While the scatter in $U$ is still rather wide, $Q$ shows a double-peaked distribution. One peak lies around 0 mJy and the other one at $\sim$0.8 mJy, the second peak is almost three times as strong as the OPC. We divided the 'Post-Burst' data set in two samples of equal size, histograms of the two samples are shown in Fig. \ref{OPCjumps}. In the first bin, the new OPC component is already present, while there is still a strong component at $\sim$ 0 mJy. In the second bin however, the new OPC component is fully dominant. We have thus detected a strengthening of the OPC directly after one of the double-peaked bursts. The component in $Q$ almost tripled it's strength.

Thus, during our monitoring campaign the OPC was stable, however, we did observe a sudden change in the OPC, directly after the second major burst in 2007. This raises the question how the OPC has evolved in the past.

A proper investigation of this subject would require long-term photopolarimetric monitoring (i.e. polarization and flux measurements) in one well-calibrated filter. Such a dataset is not available. However, to get an idea of the evolution of the OPC, we can plot the PA over time. As the frequency dependence of the position angle is weak in blazars, we can even use multi-band data. In Fig. \ref{historicPA} we plot the historic evolution of the position angle from the 1970s till the 90s. Those data are partially from literature, partially, they have been observed by Yuri Efimov with the 125 cm telescope at the Crimean Astrophysical Observatory (Ukraine), using the computer controlled \textit{UBVRI} Double Image Chopping Photopolarimeter, developed at the Helsinki University Observatory by V. Piirola. 

It is very clear from this plot that the OPC was not stable in the past $\sim$ 40 years. The most dramatic change happens around 1994, during one of the major double-peaked bursts: the preferred $PA$ shows a swing, changing its orientation from $\sim90$\degr to $\sim180$\degr. This is also the value that is observed during our monitoring campaign. An interesting finding is the fact that all strong changes in the position angle seem to have happened shortly ($\sim1$ yr) after one of the major double-peaked bursts. This is very similar to the evolution we observe during our monitoring campaign, changes in the OPC seem to follow the double-peaked bursts.

To assess this further, we present a plot similar to Fig. \ref{OPC_evo_short}, showing the evolution of the OPC. Due to the fact that the historic data is simple polarimetric data without accompanying flux measurements, we cannot calculate the Stokes parameters, thus, we will work with normalised Stokes parameters. Note that as discussed earlier in this Section, the normalised Stokes parameters are worse for determining the OPC as they are contaminated by unpolarized flux (see Fig. \ref{stokes} compared to Fig. \ref{polfluxstokes}). We present the evolution of the OPC during the last 40 years in Fig. \ref{OPC_evo_long}. We divide the data into bins using 1 January of the years 1985 and 1995. While the evolution in $U$ is rather mild, the evolution in $Q$ shows a clear bulk motion from $\sim$ -10 per cent to $\sim$ + 10 per cent during the last 40 years, the crossing of the zero point is most likely around $\sim$ 1995. This corresponds to a motion of the OPC in the Stokes plane from position angles of $\sim$ 90\degr in the 1970s to a position angle of  $\sim$ 180\degr in the present. We will not further investigate these changes as our sampling is extremely uneven and most likely also concentrated around interesting events in the light-curve. As we have seen before, Stokes plane plot look very different during quiescence or burst (Fig. \ref{OPC_evo_short}). Therefore, we believe that binning the data into smaller bins will introduce uncontrollable selection effects.

We studied statistical properties of the polarimetric dataset and found a strong alignment in position angle that originates in a strong peak in the distribution of values in the Stokes plane. We interpret this finding as a sign of an underlying source of constant polarized flux. This emission dominates during quiescent phases, while during bursts chaotic emission dominates. During our monitoring campaign, we observe a strengthening of the OPC after the 2007 burst. We also study long-term evolution in the OPC and find a rather steady migration in the Stokes plane. Also for the long term evolution, the changes in the OPC seem to correlate with the major bursts. 


\subsection{Reoccurring events in the polarimetric light-curve}
\label{sec:bubbles}

While in Section \ref{sec:double-bursts} we only discussed events that show as strong rise in total flux, in this Section we aim to determine if re-occurring events in the polarization exist. We also aim to classify those re-occurring events in a way that will make it possible for other authors to identify similar events. A table with a list of all events discussed in the following paragraphs and their basic properties is presented in Tab. \ref{bubblelist}.

To identify the 'reoccurring events' we visually inspect the 2005--2009 light-curve in $P_{Q/U}$, $Q/U$, $P$ and $PA$ and compose a list of all remarkable events. We are aware that this method is somewhat arbitrary, as the identification of the events is subjective. We are also aware that we are heavily influenced by data gaps. However, because optical polarization variability in blazars is poorly understood we believe that it can help to classify and discuss typical events. The time of all those events is also shown in Fig. \ref{alldata}, dash-dotted lines indicate the beginning and end of all events.

\begin{figure}
\includegraphics[width=8cm]{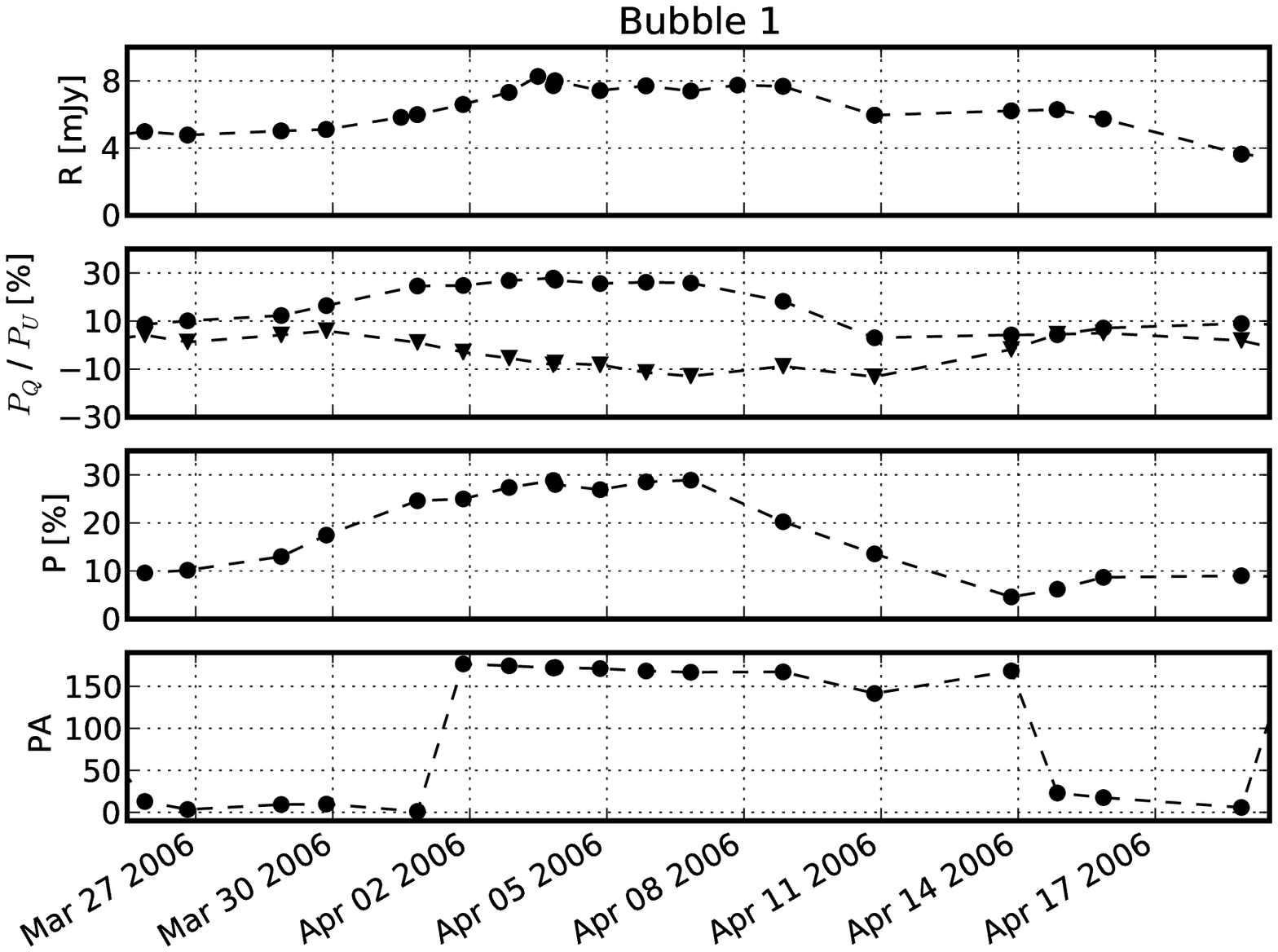}
\vspace*{0.5cm}
\includegraphics[width=8cm]{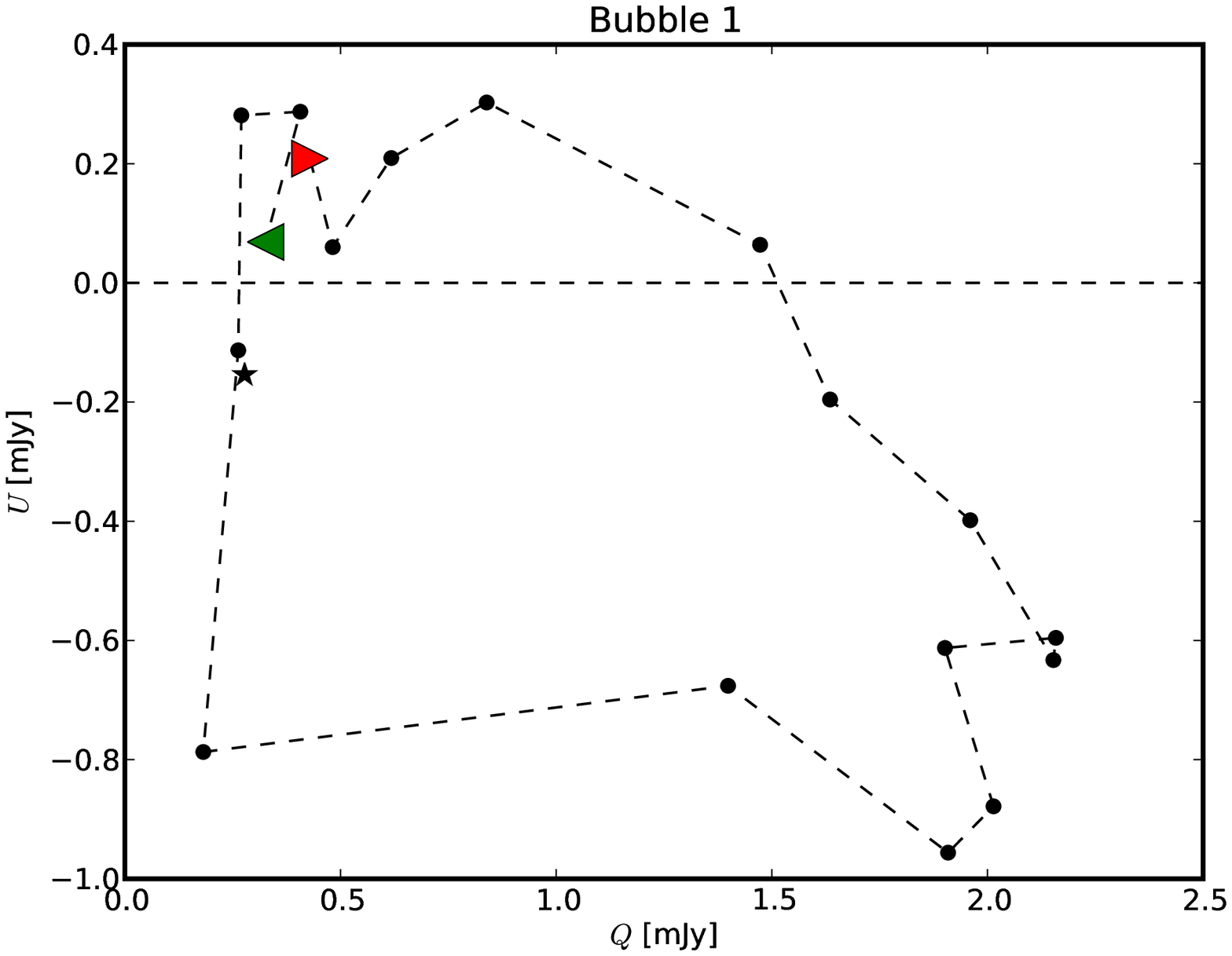}
\caption{Bubble 1: Upper plot show flux, $P_{Q/U}$ (circles: $P_{Q}$,triangles: $P_{U}$), degree of polarization and position angle, lower plot shows Stokes plane plot of the event. Red right-pointing triangles denote the beginning of the movement, green left-pointing triangles denote the end of the movement. Star symbol denotes location of the OPC.}
\label{bubble1}
\end{figure}

The first group of events with similar appearance are the 'Bubbles'. In these events, both $P_{Q}$ and $P_{U}$ start at low values, rise to a maximum simultaneously and then return to low values, enclosing a bubble. This eye-catching appearance in the plots lends this type of events the name 'Bubble'. The prototype for this group of events is Bubble 1, a 25 d lasting event that occurred between JD 2453820 and 2453845. In Fig. \ref{bubble1} we plot the flux, Stokes parameters $P_{Q/U}$, $P$ and $PA$ of the prototype Bubble 1. We see that a dramatic rise in flux, the flux almost doubles during the outburst. The normalised Stokes parameters show a bubble feature. The degree of polarization shows a rather flat flare. For the $PA$ however, we do not see any exceptional behaviour. The 'jump' from 0--180\degr is not as exceptional as it might look, 0\degr and 180\degr are identical, so what we see is merely a change by a few degrees. But while the concept of a bubble so far seems rather limited to its obvious appearance in the $P_{Q/U}$ light-curve, if we plot the bubble in the Stokes plane, it becomes clear that bubbles are more than just eye-catching features. Fig. \ref{bubble1} shows the Stokes plane plot of Bubble 1, we see a circular movement. Thus, bubbles can be used to identify circular movements in the Stokes plane. Circular motions in the Stokes plane can be interpreted as rotations of a magnetic field component. The movement in the Stokes plane in Bubble 1 is clockwise (we define the direction as seen when plotting $Q$ on the $x$--axis and $U$ on the $y$--axis). Bubble 1 has also been observed by \citet{darcangelo_synchronous_2009} within a multi-wavelength campaign. They interpreted that data as a sign for a shock front moving along the jet.

\begin{figure}
\includegraphics[width=8cm]{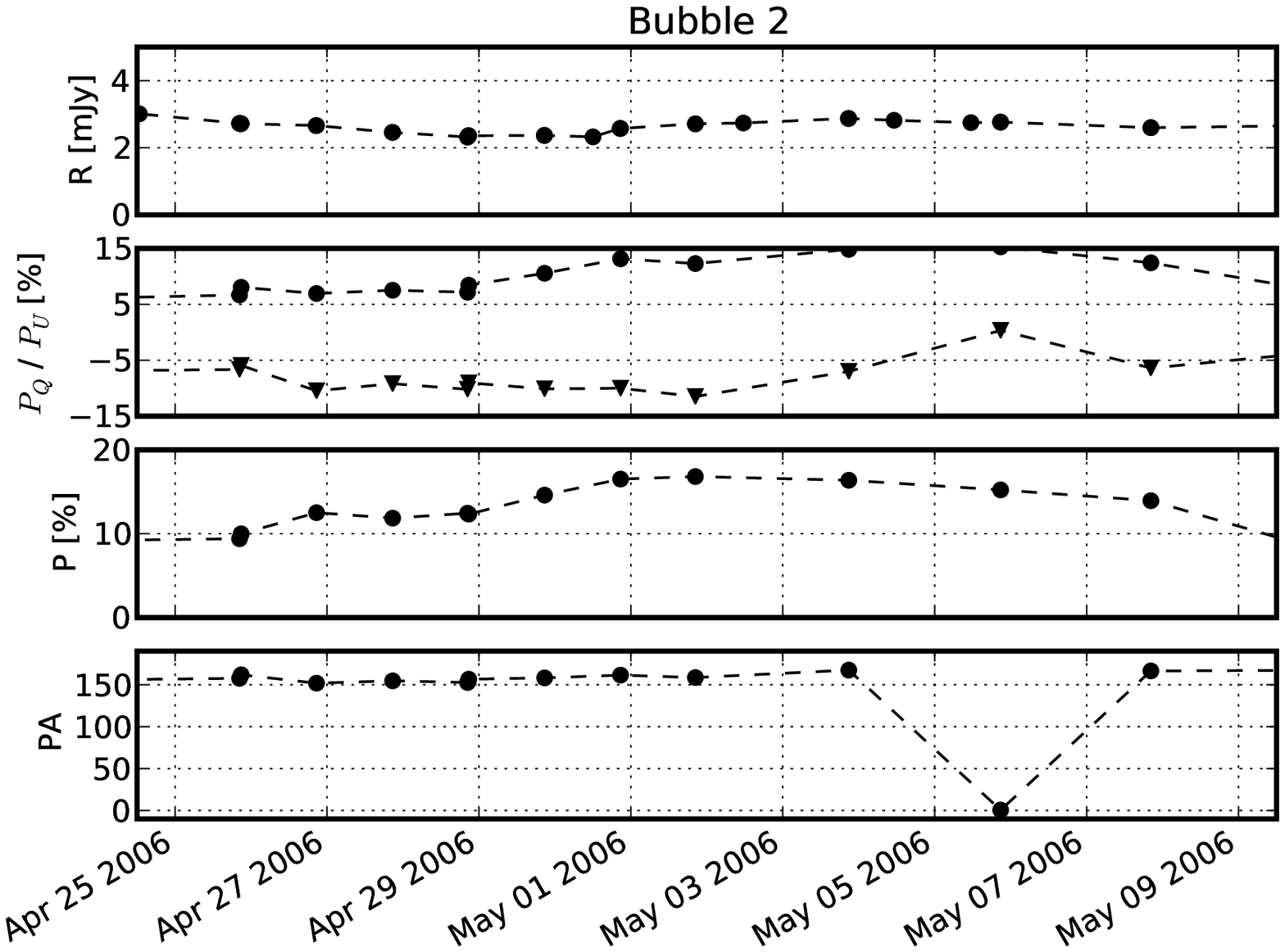}
\vspace*{0.5cm}
\includegraphics[width=8cm]{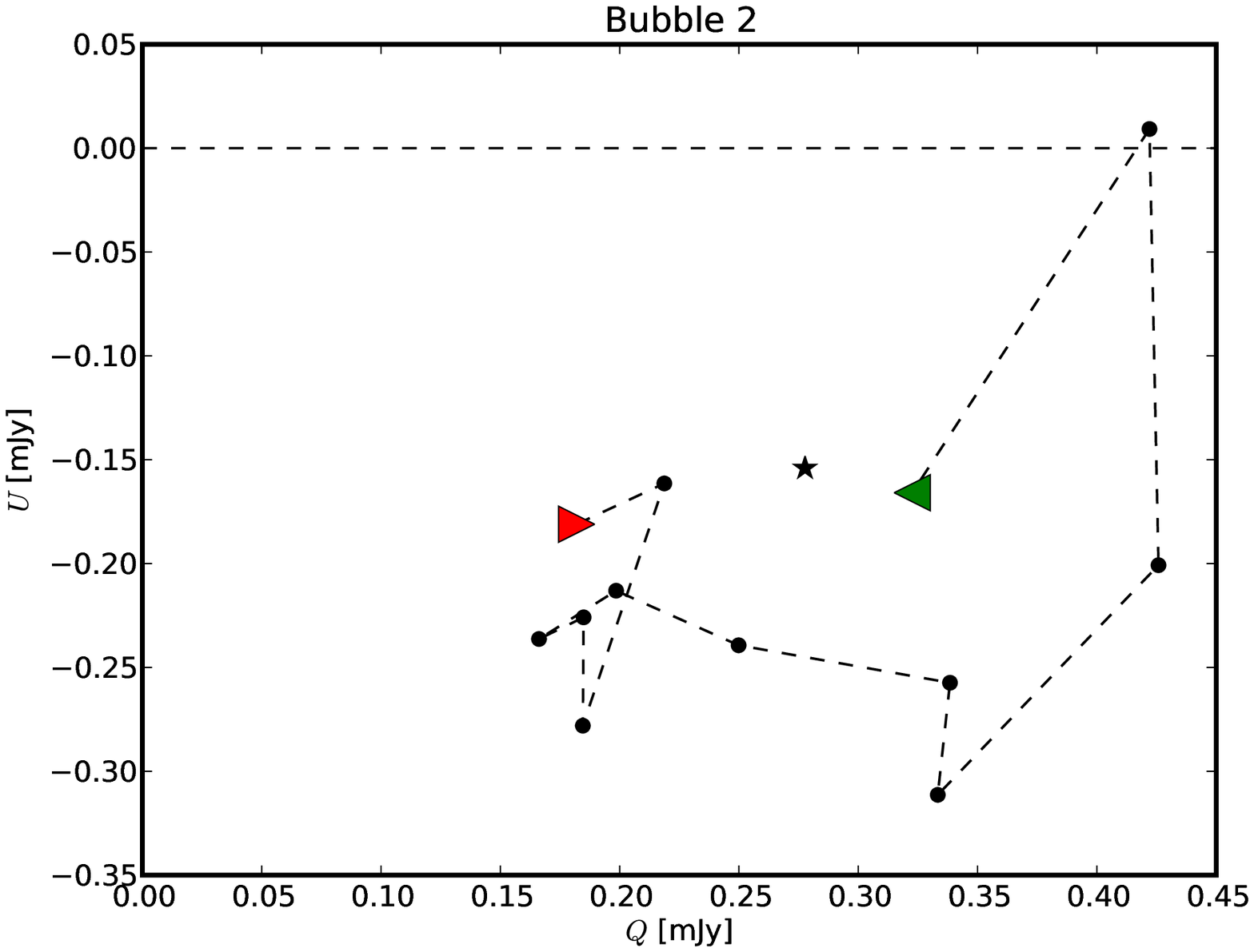}
\caption{Bubble 2, plots as for Fig. \ref{bubble1}}
\label{bubble2}
\end{figure}

We observe five Bubbles during our monitoring campaign, their properties are summarized in Table \ref{bubblelist}. Plots of those events are presented in Figures \ref{bubble1} -- \ref{bubble2}. Only one of those (Bubble 5)  is a 'false' bubble, the event looks like a bubble in the $P_{Q/U}$-plot, however, it is not a circular movement in the Stokes plane. We will discuss this event and its implications at the end of this paragraph.




The second group of events is a very special type of circular movements, those enclosing the zero point. A circular movement around the zero point shows as a swing in position angle. The position angle swings from 0\degr to 180\degr and thus once through the whole Stokes plane. We searched the light curve for these events, which we call 'Swings' and found two Swing events. Further information about those events can be found in Table \ref{bubblelist}.

\begin{figure}
\includegraphics[width=8cm]{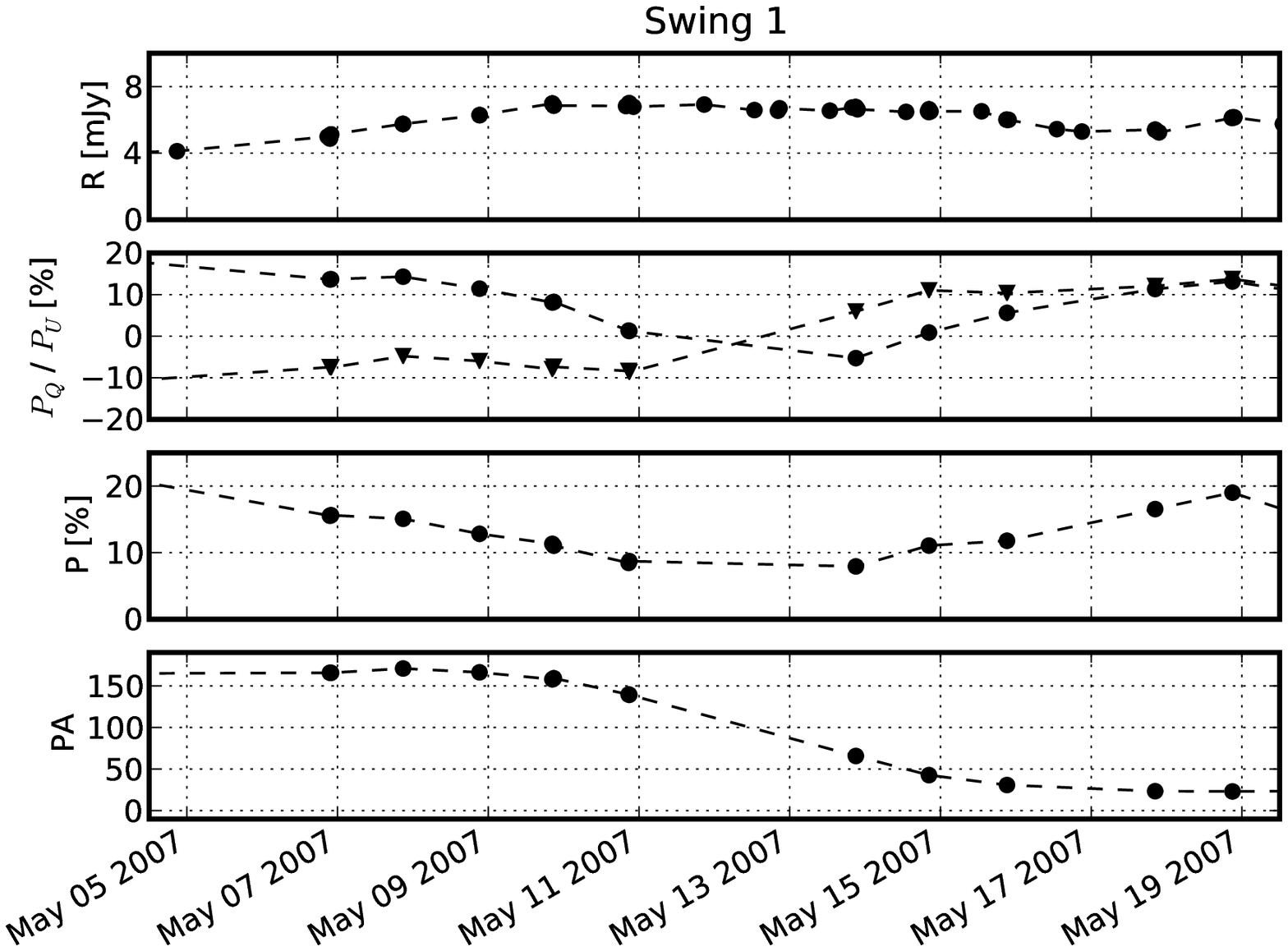}
\vspace*{0.5cm}
\includegraphics[width=8cm]{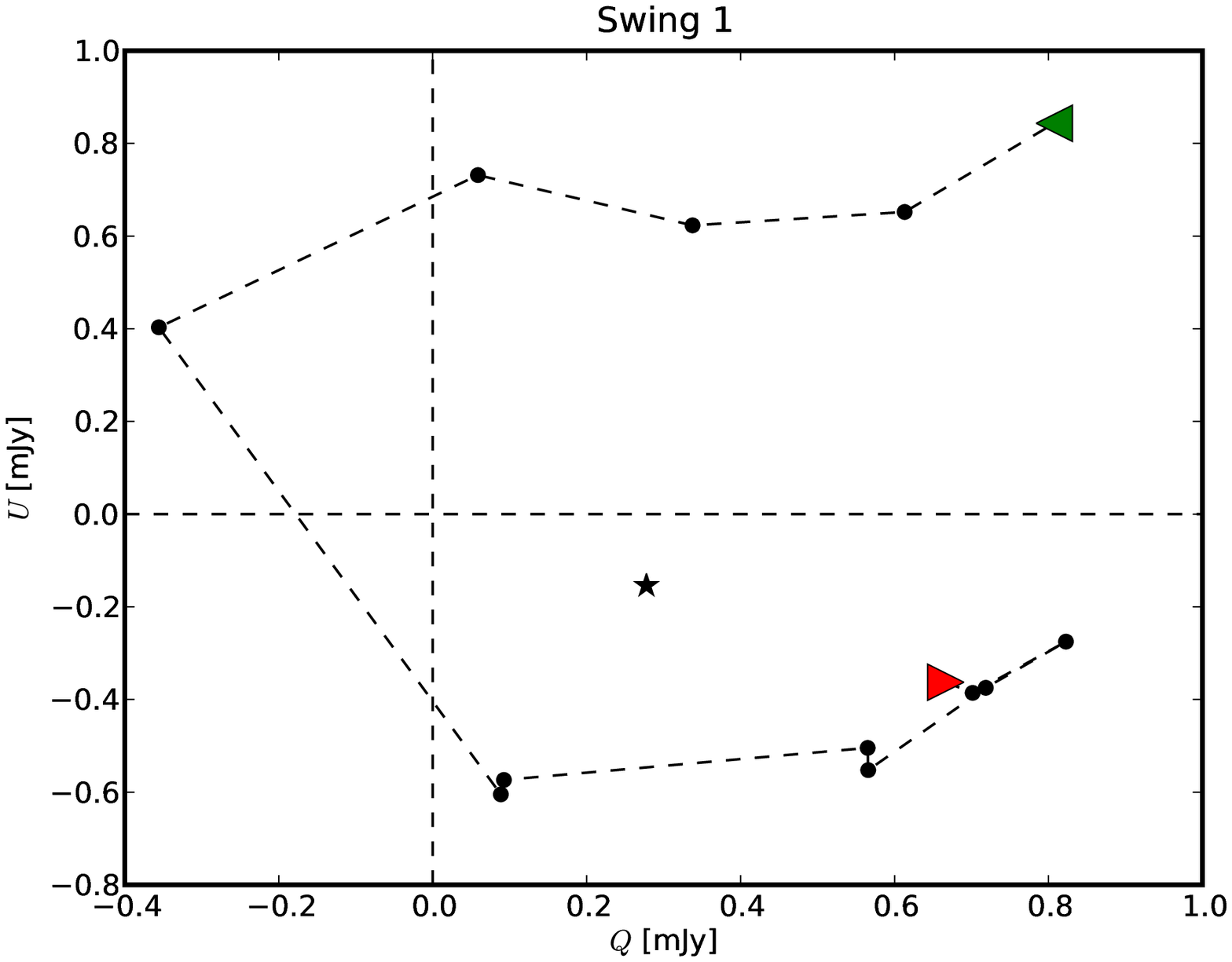}
\caption{Swing 1, plots as for Fig. \ref{bubble1}}
\label{swing1}
\end{figure}

\begin{figure}
\includegraphics[width=8cm]{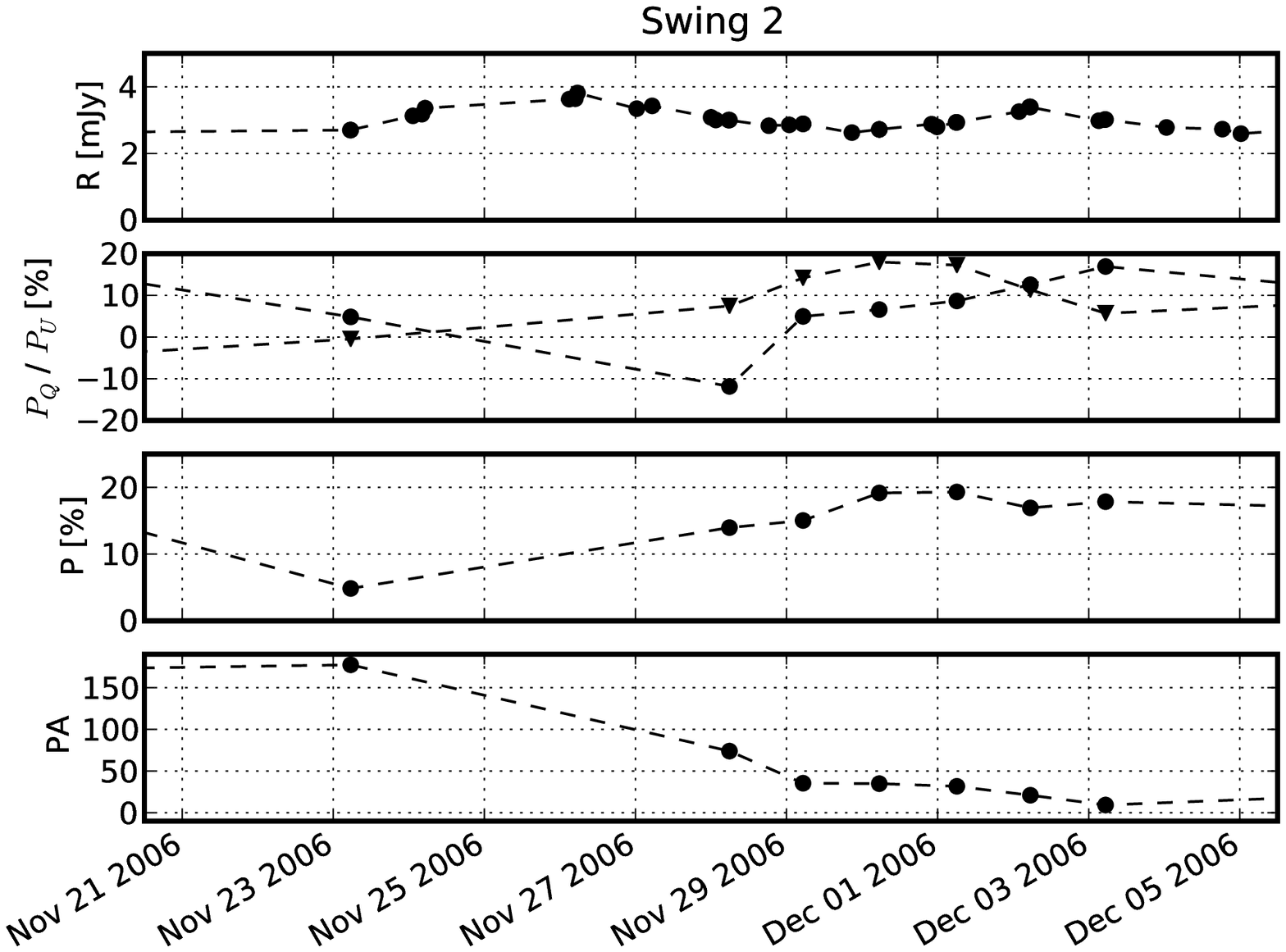}
\vspace*{0.5cm}
\includegraphics[width=8cm]{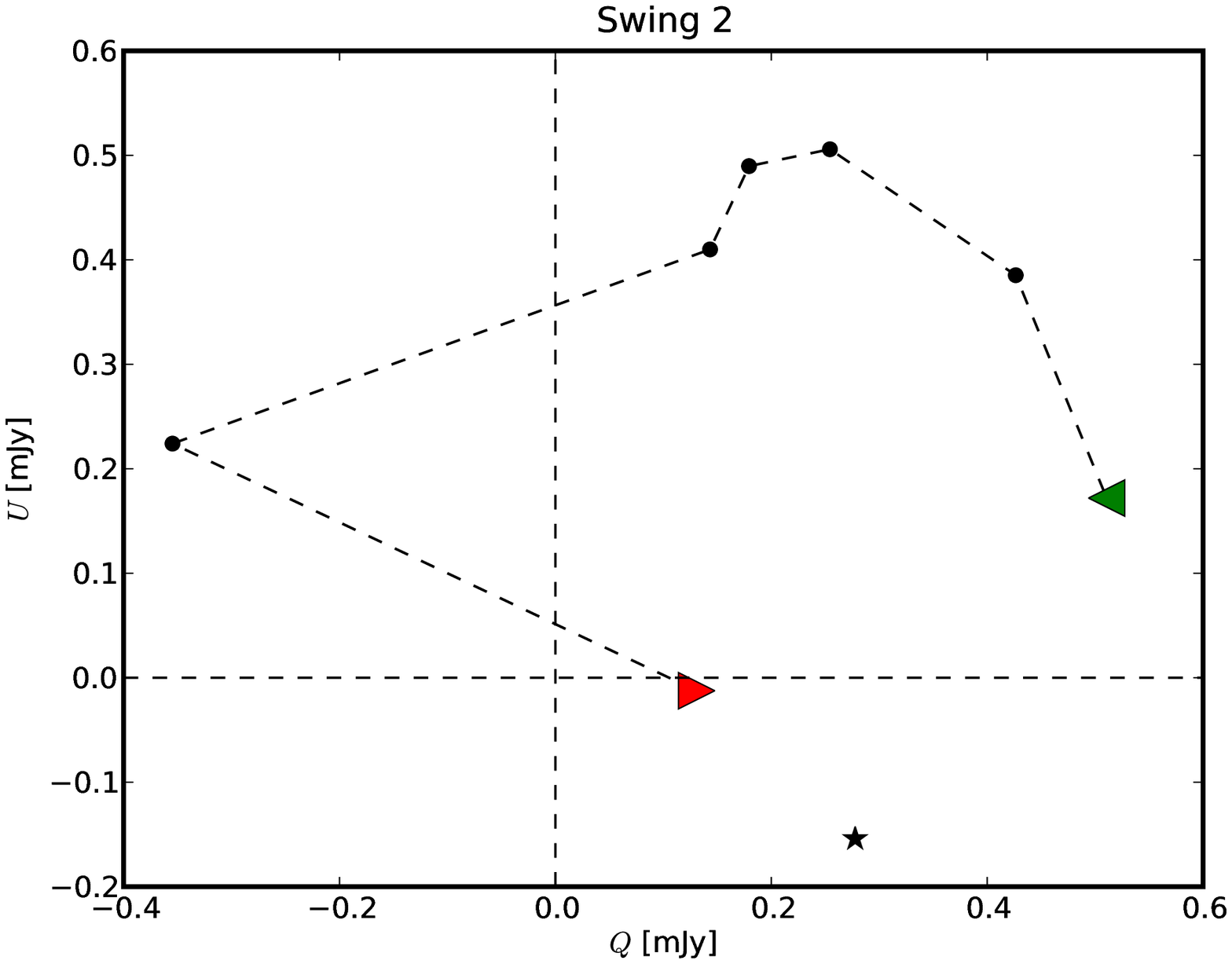}
\caption{Swing 2, plots as for Fig. \ref{bubble1}}
\label{swing2}
\end{figure}

The prototype of this class is Swing 1 (Fig. \ref{swing1}). We see a monotone evolution of the position angle from 180\textrightarrow0\degr. Opposed to the Bubbles, the degree of polarization actually decreases during the swing. In $P_{Q/U}$ we see a symmetric evolution. But due to the fact that $P_{Q}$ and $P_{U}$ are of the same sign, no bubble is enclosed. However, if we look at the Stokes plane plot, we see that Swing 1 also shows as a circular movement in the Stokes plane. One other Swing is observed: Swing 2 (Fig. \ref{swing2}).

\begin{table*}
\begin{minipage}{180mm}
\caption{List of events identified in the OJ287 2005--2009 light-curve. event ID: name of the event, will be used as ID of the event in plots and the text; JD range: JD range in start JD -- end JD, JDs are rounded to full numbers; $\Delta JD$ : length of event in days; direction: direction of movement in the Stokes plane; quadrant: quadrant in which bubble is located; circle: full or half circular movement; comments: other comments; date: human-readable date.}
\label{bubblelist}
\begin{tabular}{c|ccccccc}
\hline \hline
event ID & JD range & $\Delta JD$ [d] &  direction & quadrant & circle & comments & date\\
\hline
Bubble 1 & 2453820-2453845 & 25 & clockwise 	& lower right & full & & 25.3.2006--19.4.2006\\
Bubble 2 & 2453851-2453865 & 14 & counter-clockwise & lower right & half & & 25.4.2006--9.5.2006\\
Bubble 3 & 2453865-2453880 & 15 & clockwise & lower right & $<$ full & & 9.5.2006--24.5.2006\\
Bubble 4 & 2454360-2454404 & 44 & clockwise & lower right & $<$ full & 4 merged bubbles? & 16.9.2007--30.10.2007\\
Bubble 5 & 2454420-2454444 & 24 & ? 	    & lower right & ? & no rotation in Stokes plane & 15.11.2007--9.12.2007\\
Swing 1  & 2454227-2454239 & 12 & counter-clockwise & -- & $<$ full & & 6.5.2007--18.5.2007\\
Swing 2  & 2454063-2454073 & 10 & clockwise & -- & $<$ full & & 23.11.2006--3.12.2006\\
\hline
\end{tabular}
\end{minipage}
\end{table*}

\begin{figure}
\includegraphics[width=8cm]{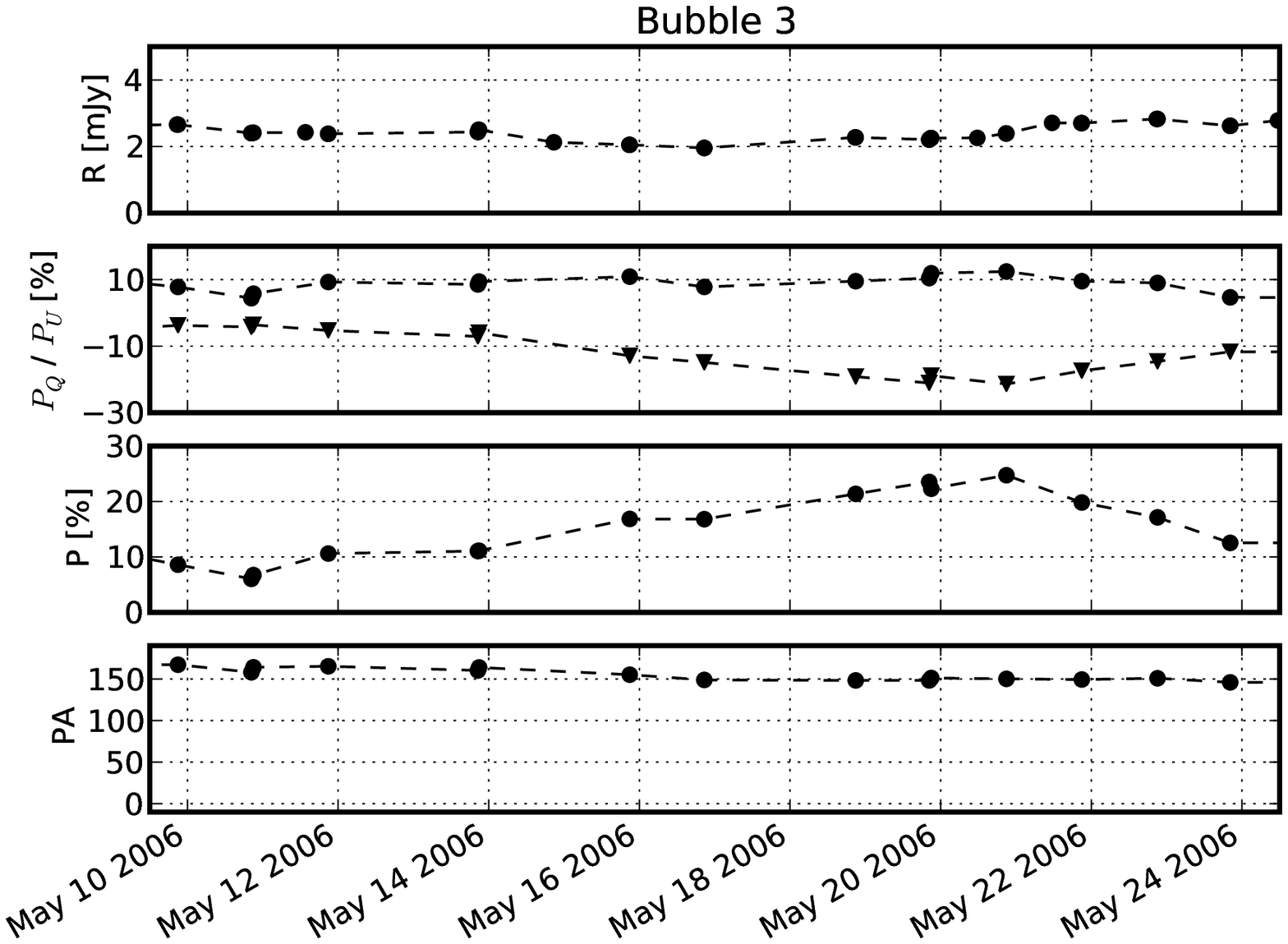}
\vspace*{0.5cm}
\includegraphics[width=8cm]{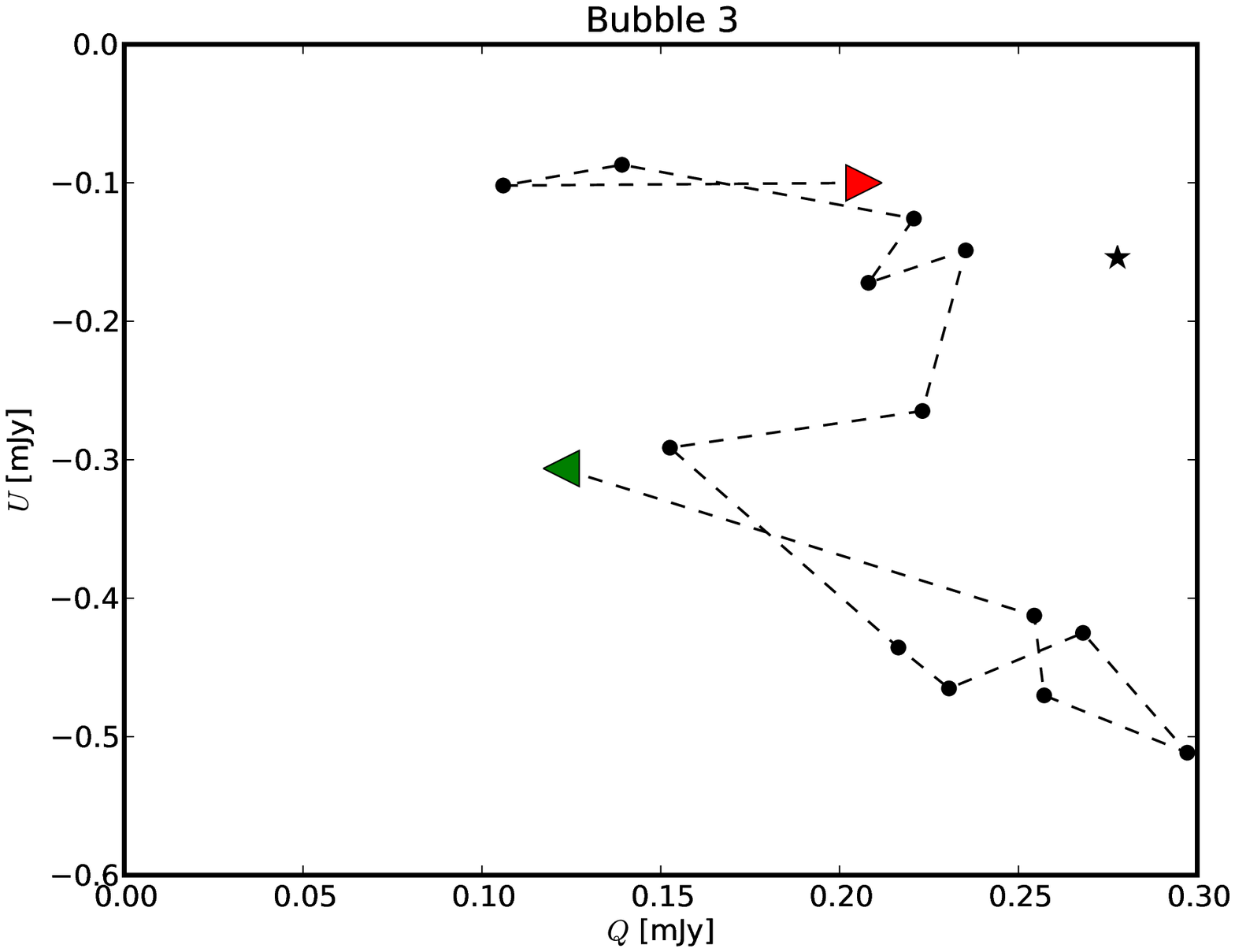}
\caption{Bubble 3, plots as for Fig. \ref{bubble1}}
\label{bubble3}
\end{figure}

\begin{figure}
\includegraphics[width=8cm]{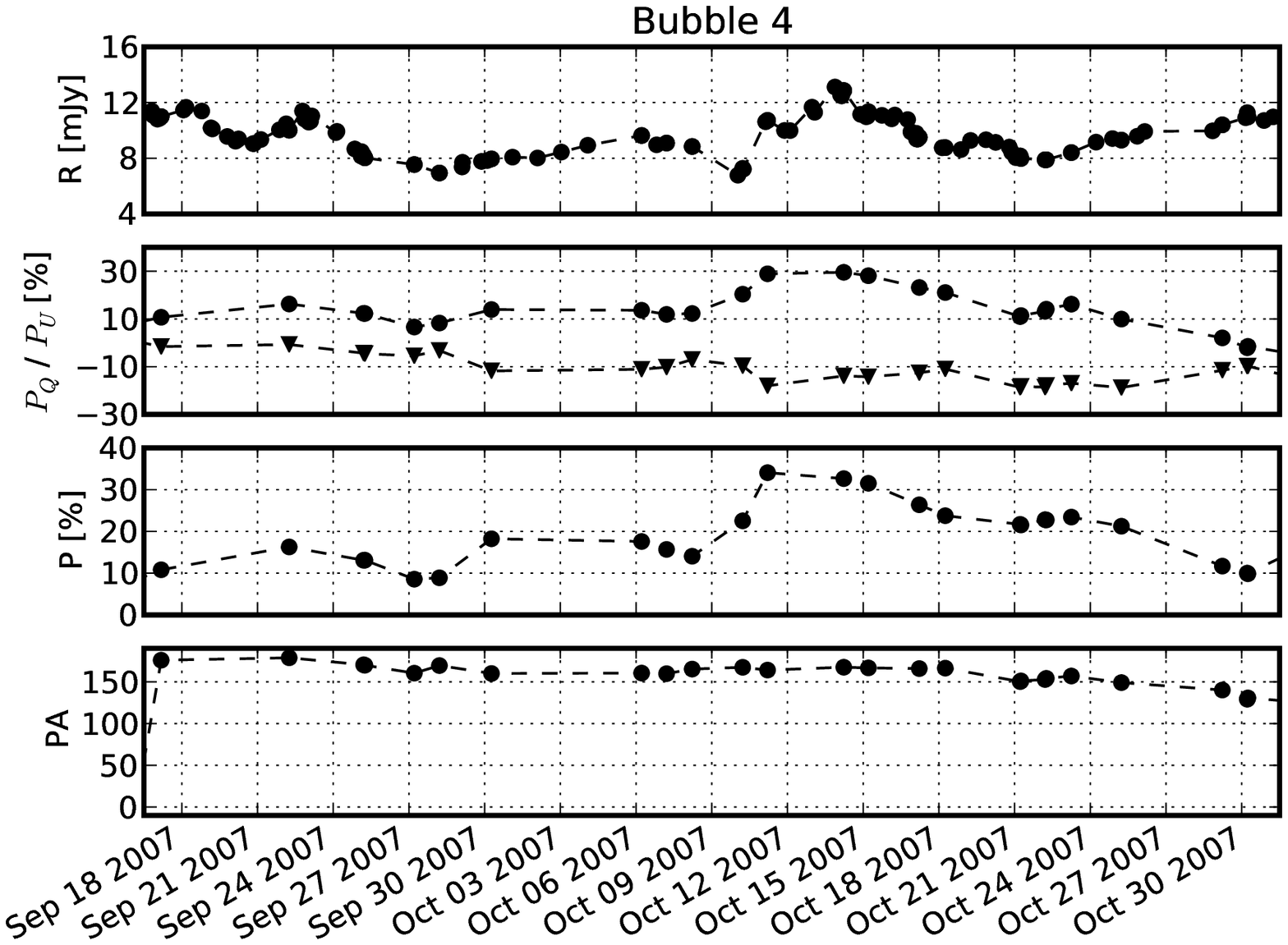}
\vspace*{0.5cm}
\includegraphics[width=8cm]{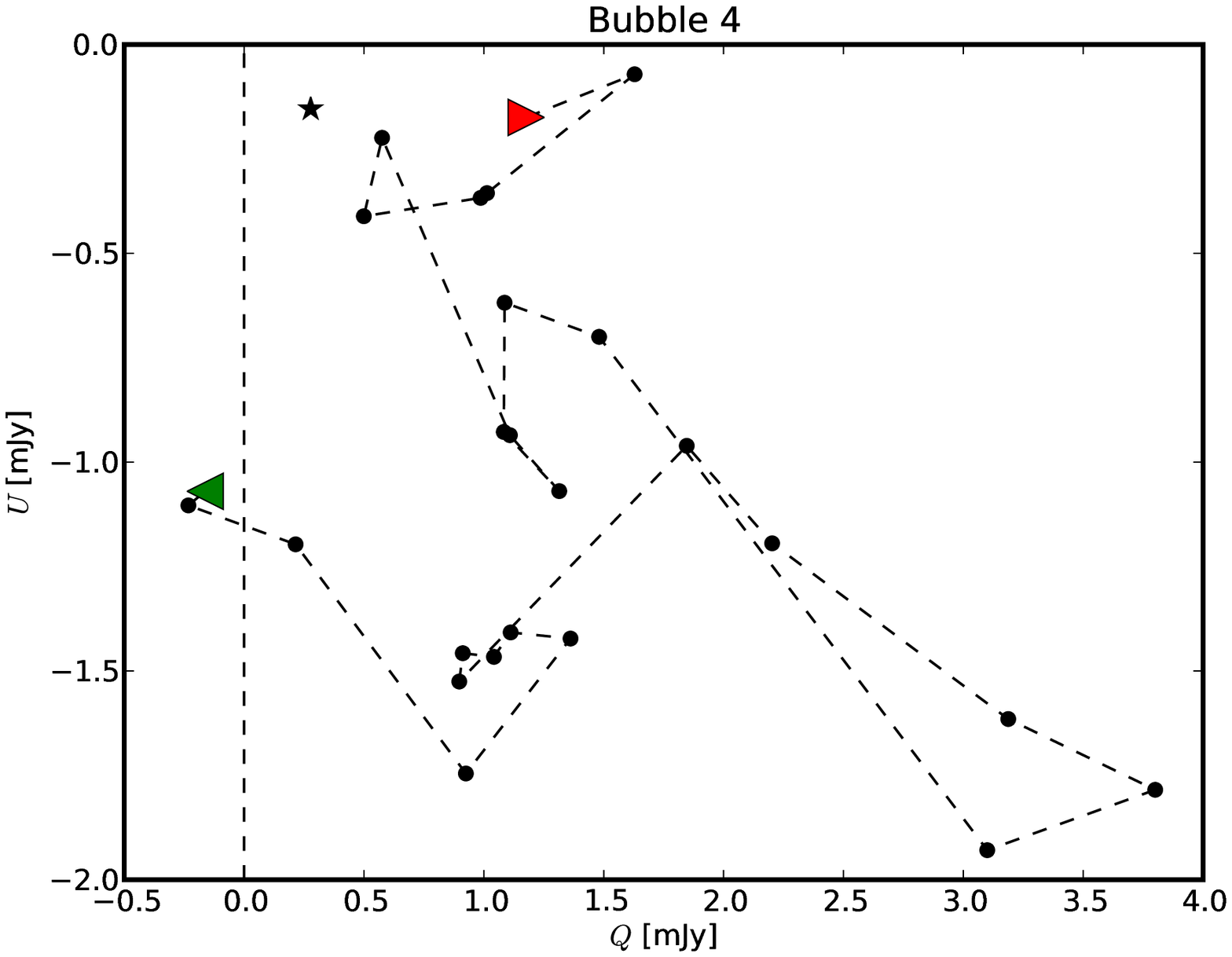}
\caption{Bubble 4, plots as for Fig. \ref{bubble1}}
\label{bubble4}
\end{figure}

\begin{figure}
\includegraphics[width=8cm]{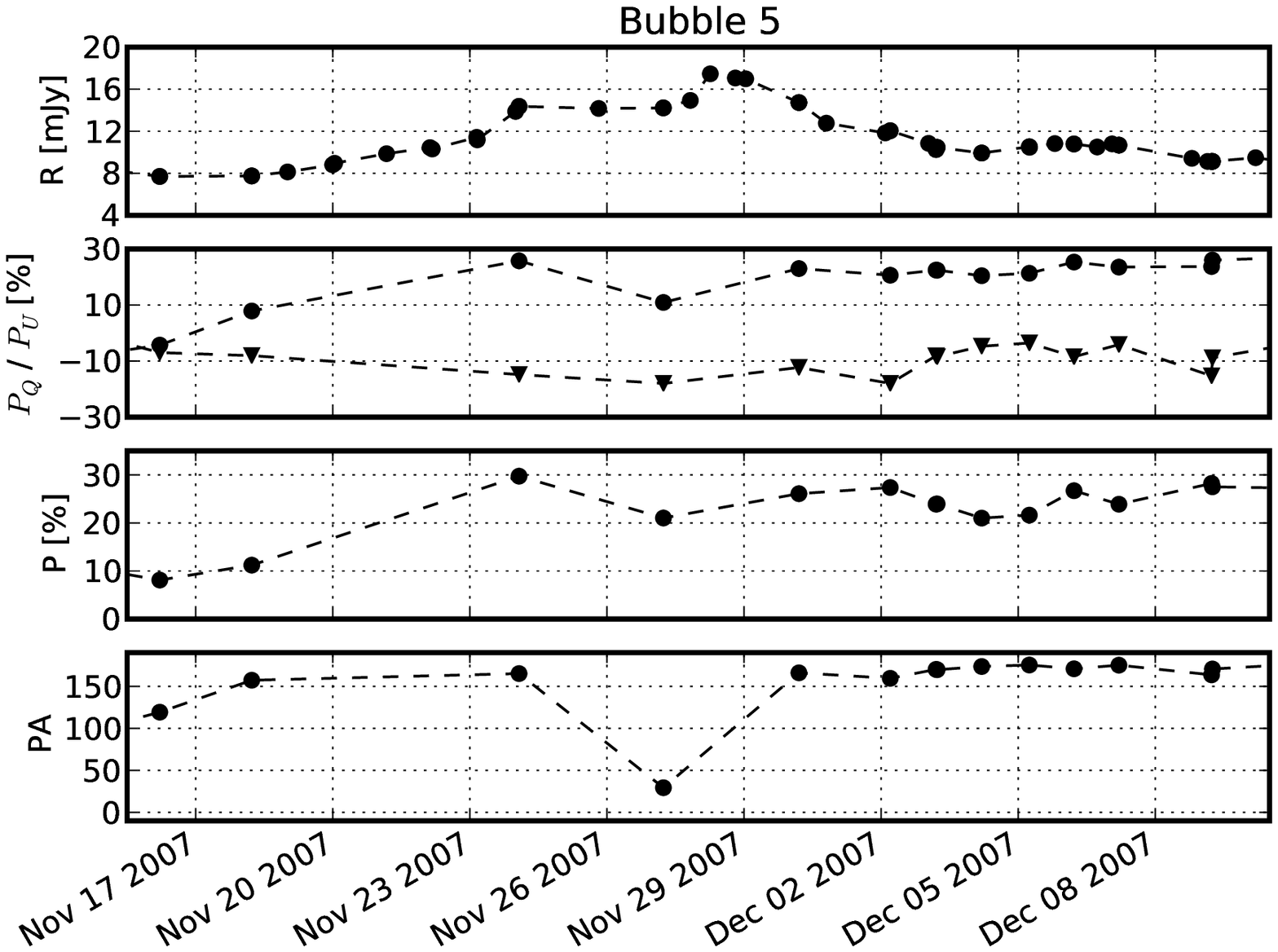}
\vspace*{0.5cm}
\includegraphics[width=8cm]{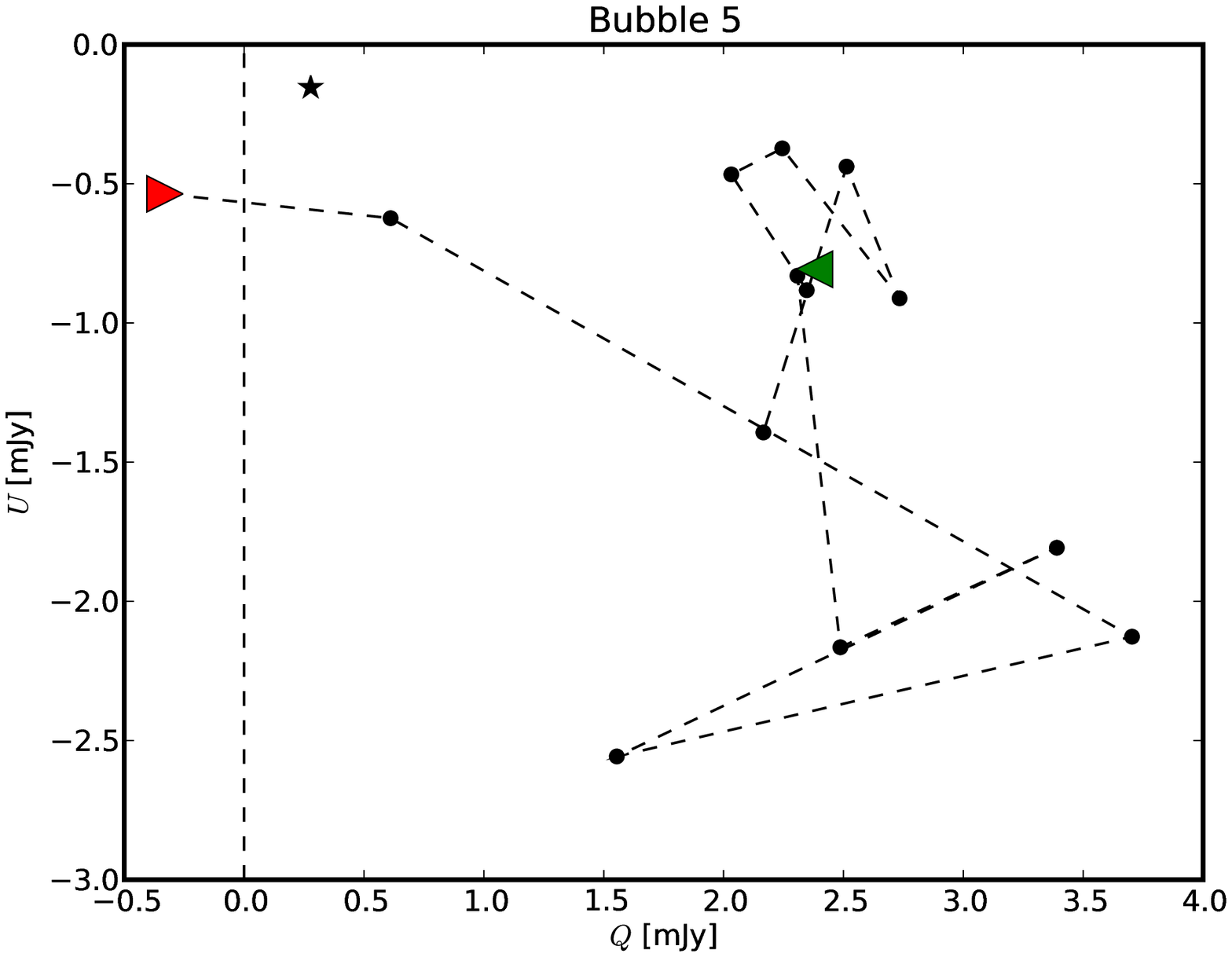}
\caption{Bubble 5, plots as for Fig. \ref{bubble1}}
\label{bubble5}
\end{figure}

As both the Swings and the Bubbles represent circular Stokes plane movements, we will from now one refer to both classes as 'rotators'. In this context it becomes clear why it is usually not sufficient to study the degree of polarization and position angle only. Polarization is a vector: while plotting the length ($P$) and orientation ($PA$) of a vector is intuitive for a single vector, it is not for a composition of several vectors. As we have shown in the previous section, the polarized emission in OJ287 can be separated into two components.

Bubbles and Swings have a completely different appearance in $P$ and $PA$, even they actually represent the same movement in the Stokes plane. Both represent circular movements, however, the fact that the centre of that movement is offset from $(0,0)$ changes their appearance dramatically. Even the size of the circle has a dramatic impact on the appearance of the circular movement in $P$ and $PA$. While a small decentred circular movement shows as a Bubble, a big circular movement will show as a Swing.

Identification of these events is therefore not straightforward. Using the circular appearance in the Stokes plane for identification is straightforward in theory. However, this method is heavily susceptible to misidentification due to irregular sampling. Rotators can be identified by their symmetrical evolution in $P_{Q/U}$ or $Q/U$, as seen both in Bubbles and in Swings. Swings can additionally be identified through their appearance in position angle. After rotator candidates have been identified using these methods, Stokes plane plots can be used to assess if the candidate represents a true rotation in the Stokes plane. For our study, six of the seven rotator candidates chosen in the above manner were confirmed by Stokes plane plots (Bubble 5 was rejected), thus this method is rather effective in identifying rotators.

Another open question is how those rotators relate to the OPC. We plotted the position of the OPC in all Stokes plane plots of the rotators (Fig. \ref{bubble1}--\ref{swing2}). For the only full circular movement (Bubble1), the OPC lies within, but at the very border of the bubble. For the others, it is not clear if the OPC lies at the centre of the movement. The OPC variability can hardly explain the offset as most rotators happen before the 2007 burst and thus the OPC was stable during those events. Thus,  rotators do not seem to have a common centre.

It has been argued that such events might simply represent random walks in the Stokes plane. \citet{jones_magnetic_1985} performed Monte Carlo simulation to address this problem. Even they only studied the probability to observe Swings (i.e. circles enclosing $(0,0)$), they found probabilities up to 30 per cent to observe such an event in a pure random walk. If we include also general rotators, the probability to observe such events by chance rises even higher. Thus, we cannot reject the hypothesis that all observed bubbles and swings are results of a random walk. However, if all rotators were indeed random events, the probability to observe clockwise or counter-clockwise movements would be equal. We observe four clockwise movements and two counter-clockwise movements. Due to the small number of events, we cannot reject the null-hypothesis that the directions of the movements are randomly distributed.

We noted before that all rotators seem to happen on similar time-scales. For this discussion we exclude Bubble 4, which might be a merging of four bubbles, and Bubble 5, which is actually not a circular movement in the Stokes plane. The only full circular movement (Bubble 1) lasts 25 d, the other circular, but not closed, movements last between 10--15 d. The finding of similar time-scales in all rotators speaks against a random walk. However, it is possible, though unlikely, that this is a selection effect. Due to the fact that the sampling density is usually $\sim$1--3 d, circular movements much shorter than 10--20 d could not be identified as they would be heavily under-sampled. As for the missing of considerable longer lasting rotator events, it is obvious that the probability to observe a certain pattern from a random walk drops for bigger numbers of data points.

While the preference of clockwise movements and the similar time-scales point towards a physical reason behind the rotators, we cannot completely reject the possibility that some of the previously described events are a result of random walks. We will however discuss possible physical causes for such observations in Section \ref{sec:discuss-general}.

\subsection{Optical spectral index variability}
\label{sec:opticalspecindex}

We investigate the variability of the spectral index $\alpha$ ($F_{\nu} \propto \nu^{-\alpha}$) in the optical by fitting power-law spectra to the observed multicolour data. The multicolour data are not strictly simultaneous, but we use only multicolour data where all filters were observed within an hour from each other.  Altogether 218 multicolour observations are used for the spectral fits.

For converting the \textit{BVRI} magnitudes to linear fluxes we used the zero points given in \citet{bessell_ubvri_1979}. The galactic absorption was corrected using \citet{schlegel_maps_1998}. The flux of the host galaxy is $\sim$ 0.08 mJy in the \textit{R}-band (see Section \ref{sec:discuss-models}), which is $\sim$ 3 per cent of the total flux at the faintest flux levels reported in this paper. Thus we have made no correction for the host galaxy.  The fits were made in a $\log \nu$ - $\log F_{\nu}$ scale by fitting a straight line to \textit{VRI} or \textit{BVRI} data using ordinary least squares (OLS) regression of y on x.  In five cases we have also \textit{U}-band data, but the \textit{U}-band point lies in all cases $\sim$ 10 per cent below the straight line delineated by other bands. Since the drop-off in the spectrum is very sudden, we treated it as a problem in the calibration rather than a
true spectral feature and excluded the \textit{U}-band from further analysis. We also excluded the three cases where the error in the spectral index was larger than 0.1.

\begin{figure}
\includegraphics[width=10cm]{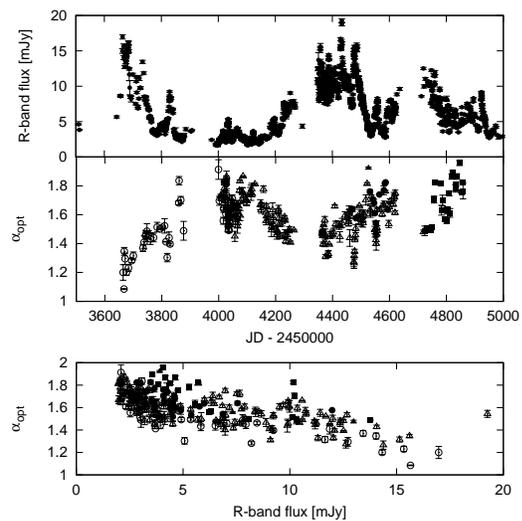}
\caption{Top panel: The R-band flux of OJ 287 between
May 2005 and June 2009. Middle panel: the optical spectral index during
the same period. Bottom panel: The dependence of the optical spectral
index on the Flux density. The different symbols refer to Osaka (open
circles), Krakow (filled circles), Suhora (open triangles), Liverpool (closed squares) and
Canakkale (filled triangles) data.}
\label{specindex}
\end{figure}

In Fig. \ref{specindex} we show the evolution of the optical spectral index with time and its dependence on the optical flux. We see a clear dependence between the flux level and the spectral index with brighter flux levels corresponding to flatter optical spectra. The Spearman correlation coefficient of the flux--$\alpha_{\rm opt}$ correlation is $-0.71$ with $>$ 99.9 per cent significance. This kind of 'bluer when brighter' behaviour has been observed before in OJ287 and also in other blazars (e.g. \citealt{fiorucci_continuum_2004}), although a more complex behaviour is sometimes observed \citep{raiteri_new_2008}. Also negative correlations between flux and bluer colours have been reported (e.g. \citealt{bttcher_whole_2009}).  During the 1993-94 outburst of OJ 287 the optical colours were also reported to have been constant over a wide range of optical magnitudes \citep{sillanpaa_double-peak_1996,hagen-thorn_variable_1998}.

If one looks at the flux - $\alpha$ plots in \citet{fiorucci_continuum_2004} it becomes evident that the spectral slope changes are very subtle and a successful detection requires monitoring the target over a large flux range and with high precision. For instance, in case of our data, the change in \textit{V}--\textit{R} colour is 0.06 mag over a magnitude range of 2 mag. Thus in a short campaign covering a limited range in flux the spectral slope changes may go unnoticed leading to apparently contradicting claims about the dependence of spectral slope on flux level.

In case of OJ287 it should be noted that while the 1993--94 seemed to exhibit no colour changes. \citet{hagen-thorn_variable_1998} noted by comparing the 1993--94 outburst to the 1983--84 and 1971--72 outbursts that the spectral index depends on the peak level of the outburst with brighter outburst exhibiting flatter optical spectra. Thus, all available evidence seems to support the view that in OJ 287 the major outbursts are accompanied by flattening of the spectrum at optical frequencies. As our data show, the flattening occurs during both major outbursts. This suggests a common origin in all major outbursts of OJ 287, including both 'sub-bursts' of the double-peaked burst. The simplest explanation is that the flattening is due to injection of high-energy electrons into the emitting region in the relativistic jet. The flattening can also be explained by two-component models, but multicolour polarimetry would be required to study this further.

\section{DISCUSSION}
\label{sec:discussion}

The discussion is divided into two parts: firstly, we discuss variability and stability in blazar jets, we will derive where the observed emission originates and what we can learn from our findings. Secondly, we discuss models proposed to explain the regularly appearing outbursts and compare their predictions with our findings.

\subsection{Variability and stability in the jet of OJ287}
\label{sec:discuss-general}

In the previous section we showed that the polarized emission from OJ287 has a clear preferred position angle. We were able to divide the optical polarized emission into two separate components: the optical polarization core (OPC) and chaotic emission with a weak alignment along the direction of the OPC. The OPC represents emission of polarized flux that is stable on time scales of years, but highly variable on time scales of decades. We observed a strengthening of the OPC during our monitoring campaign, the change happened shortly after the 2007 burst. The chaotic emission partially shows in so-called 'rotators', which represent circular movements in the Stokes plane. It is unclear if the circular movements are centred around the OPC. In this section we shall try to derive where this emission originates and what we can learn from our findings.

\subsubsection{Where does polarized emission in blazar jets originate?}

In general, polarization in blazars is caused by synchrotron emission. Synchrotron emission is observed when charged particles move in a strong magnetic field. However, to result in high degrees of polarization, the magnetic field needs to be ordered, otherwise the polarization in different directions cancels out. An obvious way to align unordered magnetic fields is through shock fronts, which are a very common feature in magneto-hydro-dynamical (MHD) jets. Shocks can compress an unordered magnetic field and produce a strong, ordered magnetic field, oriented perpendicular to the flow direction \citep{hughes_synchrotron_1989,hughes_synchrotron_1989-1,marscher_models_1985}. As shocks naturally occur in jets and produce strong, linearly polarized emission, relating the high degrees of polarization observed in blazar jets with shock fronts suggests itself.

So far, the shock front in jet model has been extremely successful in modelling flares in several blazars (e.g. \citealt{marscher_models_1985}; \citealt{marscher_inner_2008}; \citealt{darcangelo_synchronous_2009}). However, \citet{gabuzda_helical_2004} pointed out that there is evidence that points towards a global, helical magnetic field in the jet. With increasing resolution in VLBI maps, it has been found that the aligned magnetic field covers extended areas, whereas shocks are compact. Additionally, the alignment of the magnetic field has been observed to stay intact even in the presence of bends and kinks. All these observations point towards a globally -- not locally -- aligned magnetic field. Also, our finding of extremely weak correlation between high degrees of polarization and high fluxes points towards a global magnetic field. If polarized emission would only be generated in shock fronts, it should always be related to outbursts in flux, this is however not observed.

Additionally, simulations support findings of a global magnetic field. The importance of globally aligned magnetic fields for the launching and collimation of outflows has been emphasized repeatedly (see e.g. \citealt{nakamura_production_2001}, \citealt{igumenshchev_three-dimensional_2003}). In addition, helical magnetic fields are produced rather naturally: if a global magnetic field is present in the accretion disc, the accretion process will spin up the field lines, creating a helical magnetic field that can carry the jet \citep{nakamura_production_2001}.

Another interesting observation is the alignment between optical and radio polarization. \citet{gabuzda_evidence_2006} studied the relation between the optical and radio polarization in a number of blazar jets and found both values to be surprisingly well aligned. This might mean that the alignment of the magnetic field is a global phenomenon. A global magnetic field would naturally align the polarization in the same way in every region of the jet, and thus through all wavelengths. Instabilities in the jet might cause smaller regions to show a different alignment. However, the finding that radio and optical polarization are aligned does not prove the existence of a global magnetic field. A 'common point of emission' for the bulk of both the optical and radio data can also explain such a strong alignment.

\subsubsection{How can changes in the OPC be explained?}

In OJ287, the polarization vector is currently oriented perpendicular to the jet -- pointing towards a longitudinal magnetic field -- while in the past it has been oriented along the jet direction -- pointing towards a transverse magnetic field. Both the helical magnetic field and the shock front model however produce purely transverse magnetic fields. Thus, neither the helical magnetic field model nor the shock front model can explain all the observations. But before we jump to any conclusions, let us discuss what could cause an apparent flip of the magnetic field direction.

The easiest and most obvious way to flip the PA is to flip the magnetic field. However, in shock fronts, the magnetic field is by definition aligned perpendicular to the jet flow. Thus, it is not possible to flip the magnetic field in a shock front as the magnetic field takes a well-defined value. The same goes for the helical magnetic in which the toroidal component is dominant due to beaming, causing the observed magnetic field to be transverse.

\citet{lyutikov_polarization_2005} pointed out that $PA \perp \overrightarrow{B}$ is not strictly valid in AGN jets. This relation only holds in the non-relativistic case, which is a rather absurd approximation to make for AGN jets. Due to the beaming, certain components of the magnetic field become more visible in the polarization than others. However, to achieve a 90\degr flip, one would need to assume two components with a perpendicularly oriented magnetic field. While component one is fully beamed and component two is de-beamed before the flip, the beaming of the two components would need to reverse. It is very hard to imagine a setup in which such an event would take place.

Another possibility for a flip of the magnetic field would be a movement of the jet itself. In case the jet would perform some kind of swinging motion, a swinging motion of the magnetic field would be observed in polarization. Due to the fact that OJ287 has a rather small viewing angle, this might only require a movement of few degrees or less. However, it is not clear how such a movement could be achieved. Corkscrew shaped jets are observed in several sources \citep{steffen_helical_1995}, however, these jets perform a constant precessing motion on time scales of years. For OJ287 one would have to explain a single movement. With respect to the binary model, \citet{valtonen_predictingnext_2006} stated that a binary could push the disc in a certain direction at every orbit. OJ287 does not show a constantly moving jet, but a single jump. Additionally, it is questionable if a 'kick to the disc' would show as such a sudden movement in the jet. Because the magnetic fields are assumed to be the 'framing' of the jet, it is unclear if a 'kick to the disc' could be powerful enough to move the entire magnetic field within only about a year. If the jet would have undergone such a traumatic change, it should also be visible in VLBI maps. However, \citet{gabuzda_vsop_2001} published VLBI maps of OJ287 in which components with polarization vectors both transverse and longitudinal are observed, no remains of a perpendicularly oriented dead jet are visible, making the jet swing hypothesis even more unlikely.

The last option would be a change in the opacity, i.e. a change between optically thick and optically thin emission. In case of optically thin emission the observed PA lies orthogonal to the magnetic field while for optically thick emission, the observed PA lies parallel to the magnetic field. Therefore a change of the regions that dominate the emission can cause a 90\degr flip. Beaming could cause such a change. If the beaming factor changes, the restframe wavelength of the emission we observe in the optical changes. If the change between optically thick and optically thin emission is close to the restframe wavelength of the emission we observe as optical. Thus a change of the beaming factor could explain the flip. However, the transition between optically thick and thin emission usually lies in the radio frequencies (e.g. \citealt{gabuzda_vsop_2001}).

Both the helical magnetic field and the shock front model cannot produce a naturally longitudinal field and thus cannot explain the behaviour observed in OJ287. This raises the question if it is possible to produce a \textit{naturally} longitudinal magnetic field in the jet. Longitudinal magnetic fields can be produced through shear at the border of the jet that interacts with the surrounding medium, the shear-dominated area is called sheath. In this case, longitudinal fields are observed on the border of the jet and transverse fields are observed in the spine of the jet (see e.g. \citealt{gabuzda_nature_2003}). This has been observed in some sources (see e.g. \citealt{gabuzda_nature_2003}). However, in OJ287 such a spine + sheath structure is not visible. Additionally, it is questionable if the sheath could dominate over the spine for a period of time, after which the spine dominates again.

\citet{darcangelo_synchronous_2009} discussed the interesting finding that the polarization in OJ287 -- both in radio and optical -- points towards a longitudinal magnetic field. They suggested that shear aligns the magnetic field longitudinally in the core. They argued that during the times in which the field was observed to be transverse, shocks aligned the field in a transverse direction temporarily. In contrast to these findings, \citet{efimov_photopolarimetric_2002} studied optical photometric monitoring data from 1994-1997 and found evidence of a global helical magnetic field in the jet.

Poloidal magnetic fields provide a longitudinal magnetic field naturally. In simulations, injected poloidal magnetic fields are needed to produce powerful jets (see e.g. \citealt{igumenshchev_three-dimensional_2003}). However, due to the rotation in the accretion flow, the poloidal components get spun up into a helical structure. If the magnetic field structure is helical, beaming will enhance the toroidal component strongly, this will be observed as a \textit{transverse} magnetic field. If the rotation is minimal, most of the poloidal field will be preserved and only a small toroidal component will be produced. Normal accretion discs are rotating rapidly, it is thus unlikely that a standard thin accretion disc would produce a predominantly poloidal magnetic field. Radiatively inefficient accretion flows are candidates for Bondi-like and thus minimally rotating accretion flows (\citealt{igumenshchev_convection_2001}). It has been suggested that BL Lac objects accrete through such flows and not standard thin accretion discs \citep{baum_toward_1995,ghisellini_fermi_2009}. Thus it is possible that OJ287 has accreted a predominantly poloidal magnetic field.

If we assume that a poloidal magnetic field is causing the unusual position angle currently observed in OJ287, we have to explain that the magnetic field used to be oriented transversely from 1970--1994. One could argue that shock fronts dominated the emission from 1970--1994, but that seems highly constructed. It also does not explain the migration of the OPC through the Stokes plane (Fig. \ref{OPC_evo_long}). Another way to explain the observations is through the accretion of magnetic field lines. In an accretion process, not only matter, but also magnetic field accretes onto the black hole. While closed field lines can be swallowed by the black hole, open field lines cannot. When open field lines accrete, they get caught near the black hole, building a poloidal magnetic field. In case the accretion of open field lines with the same orientation continues, the poloidal magnetic field grows stronger and stronger. Finally, the poloidal component gets so strong that it dominates and 'takes over'. Thus we can explain the PA flip observed in the 1990s as accretion of poloidal field lines. In 1994, after a phase of massive accretion of both matter and magnetic field, the poloidal component got so strong that it started to dominate. The fact that the OPC strengthened after the 2007 burst also suggests that the changes of the magnetic field are related to enhanced accretion events.

\subsubsection{What is the origin of the OPC?}

The question remains where the OPC originates. Two possible explanations for an optical polarization core exist: a global magnetic field and a common point of emission. If a global magnetic field exists, polarized emission is expected from the entire jet flow. In that case, the OPC can be interpreted as a sign of the 'quiescent jet' and therefore can be used to analyse the global magnetic field in the jet. By 'quiescent jet' we mean the jet without any turbulences or shock fronts. The emission from the quiescent jet would naturally be stable as it depends only on the strength and orientation of the magnetic field, the beaming factor and the accretion rate (or more precisely, the rate at which matter flows through the jet). All of those parameters are not expected to change on small time scales. In case of a rapid change of the magnetic field, the changes propagate in the jet with the speed of light. Due to the high beaming factor in OJ287, the change might be observed to propagate with superluminal speed. The fact that changes in the OPC are observed on time scales of about a year indicates that the bulk of the OPC emission originates in a part of the jet with a size of $\gtrsim$ 1 pc.

An alternative explanation is that there is a 'common point of emission', i.e. all OPC emission originates from a small area in the jet. It is not obvious why the 'common point of emission', which is presumably extremely small, would be stable on time-scales of years. This approach can explain the alignment between the radio and optical polarization, but has trouble explaining the flip in position angle. \citet{darcangelo_synchronous_2009} argued that the position angle flip indicates a change between a normal and shear-dominated state of the jet. However, this would require a mechanism which would turn a stable 'normal' jet into a stable shear-dominated jet.

As argued before, the OPC swing can most naturally be explained assuming a global magnetic field. The stability of the OPC on time-scales is also more naturally explained by an extended source of emission. Therefore, we argue that the OPC traces the quiescent jet and can therefore be used to study the jet magnetic field in blazars.

\subsubsection{Is the OPC commonly observed in blazars?}

Another question is if a preferred position angle, as observed in OJ287, is a common feature in all blazars. \citet{jones_magnetic_1985} observed 20 compact radio sources in different radio bands and found a 'radio polarization core', which they described as scattering around a common value. Opposed to our findings, \citet{jones_magnetic_1985} concluded that the $P_{Q/U}$ gave a better estimate of this value than $Q/U$. They concluded from this that the magnetic field is mostly tangled, only showing minor anisotropies.  \citet{angel_optical_1980} were the first ones to address this topic in the optical. They were able to divide a sample of 21 objects (all blazars or Seyfert galaxies with blazar-like SEDs) into two distinct groups. Group one does not show any preferred PA, the data-points are 'all over the compass'. We will call these sources OPC-weak in the further discussion. The second group consists of objects in which repeated measurements show a restricted range of angles. These objects will be referred to as OPC-strong in the further discussion. \citet{angel_optical_1980} already noted that OJ287 belongs to both classes, depending on the time when it is observed. During the 2005--2009 OJ287 was clearly OPC-strong, it was also OPC-strong during the monitoring campaign conducted by \citet{hagen-thorn_oj_1980}. In the 1980s however, OJ287 was OPC-weak. \citet{angel_optical_1980} noted that all OPC-weak sources in their sample are rather bright, high-redshift objects, while the OPC-strong sources are at lower redshift and fainter.

We decided to compare the both groups, omitting OJ287. We found that the two groups, the OPC-weak and OPC-strong sources, show significant differences. Out of the eight OPC-weak sources, five are FSRQs, two are BL Lacs and one (3C279) is a BL Lac/FSRQ transition object. On the other hand, out of the 12 OPC-strong sources, nine are BL Lacs, two are Seyfert galaxies and only one is a FSRQ. Even considering the fact that we are clearly limited by low number statistics and some objects might have been misidentified due to limited amount of available polarization data, this is a truly fascinating finding. If this neat separation can be confirmed for a larger sample, this would mean that that the jet emission in BL Lac objects and FSRQs differs significantly. This is supported by the findings of \citet{jannuzi_optical_1994} which studied optical polarization variability in BL Lacs and found all of them to have a preferred position angle, showing in a similar manner as observed in OJ287.

This result indicates a difference in the ratio of turbulent (unaligned) and quiescent (OPC/aligned) jet emission. In FSRQs, the turbulent jet emission seems to dominate, the OPC emission is weak compared to the normal level of the turbulent emission. The position angle is either unaligned or only weakly aligned. In BL Lac objects, the OPC emission is strong compared to the levels of the turbulent emission, the position angle is aligned. Assuming that the OPC emission originates from the common point of emission, this would indicate that the common point of emission (e.g. a standing shock front) in FSRQs is weaker compared to the rest of the jet emission than in BL Lacs. Assuming that the OPC is caused by quiescent jet emission in presence of a strong helical magnetic field, this would indicate that the magnetic field in FSRQ jets might be weaker or that the jet emission is dominated by turbulent unaligned emission. Thus, if the findings from \citet{angel_optical_1980} would be confirmed for larger samples, this would indicate that not only do FSRQs and BL Lacs show different accretion processes, but they also differ in the variability and properties of their jets.

\subsubsection{How can 'rotators' be explained?}

Another open question is the origin of the rotators we observed. Rotators are commonly observed in radio, \citet{aller_polarization_1981} first identified these events calling them 'polarization rotators'. \citet{jones_magnetic_1985} observed such rotators as well, however, he suggested that those might just be random walk events. As we discussed in Section 
\ref{sec:bubbles}, we cannot clearly reject the hypothesis that those rotator events are indeed random events.

While one might expect that all rotators would be centred around the OPC, this cannot be confirmed for our observations. The decentring of the bubbles cannot be explained assuming fast variability in the OPC. This might either be explained by another component that is constant during the rotator and decentres the rotator from the OPC or by assuming that rotator are not rotating around the zero-point but another point in the Stokes plane.

Rotators could be interpreted as blobs moving along a helical trajectory inside the jet. However, there is a serious caveat to this interpretation. If the rotators were indeed signs of a blob moving along a helical trajectory, all of them should have the same direction as the orientation of the helical field will decide the orientation of the rotation. Clearly not all rotators have the same direction! One could argue that the rotators with the 'wrong orientation' are random movements while the ones with the 'right orientation' are blobs moving along the magnetic field. However, this is clearly not a convincing explanation.

A more natural explanation is to assume that the rotators are shock fronts moving backwards and forwards in the jet, swiping through the helical magnetic field. This would explain the different rotation directions observed as well as the presence of rotators that do not represent full circles in the Stokes plane. Additionally, the 'wobbling shock front' seems to fit the bubble sequence consisting of Bubbles 1, 2 and 3. These bubbles follow each other with a very small time lag. Bubble 1 is clockwise, 2 counter-clockwise and 3 clockwise again, this could be explained as a shock front wobbling back and forth. It can also explain a common time-scale for all rotator events in case the speed of the 'wobbling shock front' is approximately constant.
\\

We are able to interpret our observations as a sign of a jet that has two major components, the quiescent jet and turbulent emission from wobbling shock fronts inside the jet. The quiescent jet contributes considerably to the jet emission, even if we assume that the emission of the quiescent jet is maximally polarized (which is rather unlikely), we find that in contributes $\sim$20 per cent to the flux in a quiescent state. We further argue that polarimetric monitoring can be used to study the global magnetic field in the jet and assess if the jet is dominated by turbulent or quiescent emission. We also find differences between FSRQs and BL Lacs concerning the strength of the quiescent jet component, our findings indicate that FSRQ jets are dominated by turbulent emission while BL Lac jets are dominated by quiescent jet emission. Further observations will have to show if our findings hold for bigger samples.  

\subsection{OJ287: a blazar hosting a binary black hole?}
\label{sec:discuss-models}

This part of the discussion is dedicated to the mystery of the regularly appearing double-peaked outbursts observed in OJ287. Before we rate the different models proposed for OJ287, we compose a list of requirements for a successful model and use this list to systematically assess each particular model.

\subsubsection{A list of requirements for OJ287 models}
\label{sec:OJ287list}

While the light-curve of OJ287 has been studied excessively, limitations to the suggested models set by the objects itself have received little attention. It is clear that a successfully model would not only have to describe the light-curve correctly but also be consistent with the properties of the object, including the black hole mass. Before discussing prominent features of the light-curve, we therefore discuss the limitations set by the fact that OJ287 is a BL Lac object and derive an estimate for the black hole mass.

BL Lacs are a special subclass of blazars. It is nowadays widely believed that BL Lac objects are Fanaroff-Riley I (FR I) type objects with a jet pointed directly towards the observer \citep{baum_toward_1995}. FR I objects are low-luminosity radio galaxies with a jet that is luminous and relativistic near the core, while it decelerates to sub-relativistic speed on scales of $\sim100-1000$ pc.

\citet{baum_toward_1995} systematically studied the differences between FR II and FR I radio galaxies and found that the spectrum of FR Is is missing strong broad emission lines. They found the luminosity and ionization of narrow emission lines is consistent with being caused by ionizing radiation from the host galaxy and not the AGN. The weakness or absence of broad emission lines cannot be explained using obscuration. \citet{baum_toward_1995} were able to explain the line properties by assuming that FR Is accrete through radiatively inefficient accretion flows (RIAFS). Therefore the ionizing radiation from the accretion disc in FR Is is missing, causing the weakness or absence of broad and narrow emission lines. \citet{ghisellini_fermi_2009} studied $\gamma$-ray properties of a big sample of BL Lacs and FSRQs and found a neat separation of the two blazar types in the $\gamma$-ray luminosity vs spectral index plane. They were able to explain this separation assuming that FSRQs accrete radiatively efficiently while BL Lacs accrete radiatively inefficiently.

\citet{baum_toward_1995} and \citet{ghisellini_fermi_2009} based their studies on completely different datasets, but reached the same conclusions. Additionally, no other possible explanation for the dichotomy between FR I / II radio galaxies and BL Lacs / FSRQs exist to our knowledge. Thus, we can state that BL Lac objects most likely accrete through RIAFs. This statement raises the question if accretion through a RIAF instead of a standard thin accretion disc would have any influence on the suggested models for OJ287.

Radiatively inefficient flows can appear in two regimes of accretion: sub-Eddington accretion ($\dot{m}\ll1$) and super-Eddington accretion ($\dot{m}\gg1$). Both flows have a common feature: due to the fact that the flows are not radiating efficiently, the cooling is inefficient. Thus, these flows are hot and geometrically thick. Over the years, several analytical models for RIAFs were developed: such as the advection-dominated accretion flows (ADAF), adiabatic inflow-outflow solution (ADIOS), the convection-dominated accretion flow (CDAF) and convection dominated Bondi-flows (CDBF). For a full description of the analytical models see \citet{igumenshchev_radiatively_2004}. In the last years, numerical viscous and MHD models were used to simulate radiatively inefficient accretion flows (e.g. \citealt{igumenshchev_three-dimensional_2003}). From MHD simulations it became clear that the properties of the flow depend strongly on the injected magnetic field. Even the exact physics in these flows is not well understood yet, one point is clear, RIAFs are geometrically thick! The lack of cooling 'puffs up' the disc. A thick disc might be problematic for binary black hole models that associate the bursts with disc crossing. In case of a thin disc, the time it takes the secondary black hole to cross the disc is short. In case of a geometrically thick disc and a small orbit however, it might take the secondary a considerable percentage of the orbital period to cross the disc. Thus the burst would cover a longer timespan.

Therefore, we estimate if the assumed trajectory of the secondary black hole lies within the simulation regime and thus within the range in which the flow is geometrically thick. The simulations from \citet{igumenshchev_three-dimensional_2003} reach out to a radius of 256 $R_{Schwarzschild}$. Considering two example masses for the central supermassive black hole of $10^{9}/10^{10} M_{\sun}$ this corresponds to $\sim 8 \times 10^{14/15}$ m. Using the third Kepler law, we can roughly estimate the semi-major axes of a possible binary black hole. We assume an orbital period of $9$ yr (in the restframe of the object). For the primary black hole mass, we use the two example values of $10^{9}/10^{10} M_{\sun}$. We furthermore assume that $M_{secondary BH} \lll M_{primary BH}$. This yields a semi-major axes for the $10^{9} M_{\sun}$ black hole of $\sim 6 \times 10^{14}$ m, for the $10^{10} M_{\sim}$ black hole, we get a semi-major axes of $\sim 1 \times 10^{15}$ m. These values lie inside, but in the outer part of the simulation box. Considering the newest version of the Lehto \& Valtonen model, in which the secondary orbit is highly eccentric, the radii of the disc crossing lie at $10^{14-15}$ m and thus well within the simulation box.

Assuming that OJ287 is indeed powered by a radiatively inefficient, geometrically thick accretion flow, the outburst due to piercing of the accretion disc by the secondary black hole lasts longer. A clearly defined 'disc piercing' no longer exists as the secondary is inside the disc for a considerable part of the orbit. However, OJ287 is only observed to be 'in burst' about ten percent of its time! In case of the Lehto \& Valtonen model the flare associated with the disc piercing lasts only $\sim$1 week! This corresponds to only $<$1 per cent of the orbital period and is thus in clear contradiction to a geometrically thick accretion flow.

Further limitations to possible models can be set by the mass of the central black hole. Due to the fact that OJ287 is a BL Lac object and does not show broad emission lines except in its extremely faint state (see e.g. \citealt{sitko_continuum_1985}), determination of the mass of the central black hole is not feasible using reverberation mapping. Another method to determine the masses of central black holes is using relations between the host galaxy and the mass of the central black hole (see \citealt{novak_correlations_2006} for a summary of these methods), this method can be used for OJ287 as the host galaxy properties have been studied. We summarize different results of host galaxy studies in Tab. \ref{host}. We use the results to estimate the mass of the central black hole.

For all publications apart from \citet{hutchings_optical_1984}, where no apparent magnitudes are given, we calculate the absolute magnitude applying k-corrections from \citet{bicker_chemically_2004}. No evolutionary corrections are applied. If the preferred morphology is a disc model, we use k-corrections for a Sb galaxy, otherwise we use the k-corrections for an E galaxy. \citet{odowd_host_2005} used the $STIS$ filter \textit{F28x50LP} for which no k-corrections are available. This filter is a low pass filter with a cut-off frequency of $\sim$5500\AA{}. In this case, we use k-corrections for Bessel \textit{R}. Even the real transmission curve of this filter is much bluer than for a Bessel \textit{R}, we do not expect that this introduces big errors. For example the difference between the k-correction for \textit{R} and \textit{I} at this redshift are only $<$0.1 mag.

\begin{table*}
\begin{minipage}{160mm}
\caption{Properties of OJ287 host galaxy from different publications. Columns have following meanings: first column shows the publication used; band: filter used; $m$ : apparent host magnitude; $M$: absolute host magnitude; $r_{e} [kpc]$: effective radius of galaxy in kpc; morph.: morphology, if not specified, none given in the publication; dc?: decentring of host with respect to point source (yes, no, ? if not mentioned, values given in arcseconds); $log M_{BH}$: log of the mass of the central black hole in solar masses. }
\label{host}
\begin{tabular}{lccccccc}
\hline \hline
 & band & $m$ & $M$ & $r_{e} [kpc]$ & morph. & dc? & $log M_{BH}$\\
\hline
\citet{hutchings_optical_1984}	& \textit{R}        & --             &  $>$-23.1 & --     & --            & ? & $<$9.0 \\
\citet{yanny_hubble_1997}	& \textit{F814W}    & 18.30$\pm$1.30 & -23.05    & --     &  --           & $~0.4$\textacutedbl & 8.9 \\
\citet{wright_host_1998}	& \textit{K}        & 12.0           & -29.15    & 6.68   & disc          & ? & --\\
\citet{wright_host_1998}	& \textit{K}        & 12.2           & -29.14    &  11.13 & none preferred& ? & 10.4 \\
\citet{heidt_observations_1999}	& \textit{R}        & 18.41          & -22.96    & 4.4    & bulge         & yes & 8.9 \\
\citet{heidt_observations_1999}	& \textit{R}        & 19.32          & -22.00    & 10.2   & disc          & yes & -- \\
\citet{urry_hubble_2000}	& \textit{F702W}    & $>$18.53       & $>$-23.01 & --     & --            & ? & $<$8.9 \\
\citet{pursimo_deep_2002}	& \textit{R} 	    & 18.91	     &  -22.79   & 4.5    & bulge         & ? & 8.8 \\ 
\citet{nilsson_r-band_2003}	& \textit{R}        & $>$18.1        & $>$-23.27 & --     & --            & yes & $<$9.0 \\
\citet{odowd_host_2005}		& \textit{F28x50LP} & $>$18.53       & $>$-23.02 & --     & --            & ? & $<$8.9 \\
\hline
\end{tabular}
\end{minipage}
\end{table*}


A few things are notable when talking about the host galaxy of OJ287. The first is the repeatedly mentioned decentring of the extended emission \citep{yanny_hubble_1997,heidt_observations_1999,nilsson_r-band_2003}. The only mentioning of the amount of decentring is from \citet{yanny_hubble_1997}, who found a value of $~0.4$ arcsec, corresponding to as much as 1.81 kpc. However, \citet{yanny_hubble_1997} might have had problems with the PSF and thus these results should be taken with caution. Decentring might be more of a sign that the host was actually not resolved than a sign that there is actually a truly decentred nebulosity. Another noteworthy fact is the extreme optical-NIR colour indicated by the different results. The measured host galaxy in \textit{K} is about 3 mag more luminous than a $L^{*}$-galaxy while the upper limits for the optical filter are around the values for $L^{*}$-galaxies. This implies an extremely red galaxy. Even ULIRGs, supposedly the reddest galaxies in the optical-NIR range, do not show such extreme optical-NIR colours but a rather flat spectrum (see e.g. \citealt{kim_optical_1995}). This leads us to assume that the detection in NIR is most likely spurious, as already suspected by the authors \citep{wright_host_1998}. All in all, we think that it is safe to assume that the host galaxy of OJ287 is unresolved, however, several consistent and reliable upper limits exist. The values found in the optical are well consistent with an $L^{*}$ elliptical galaxy. Thus we can use these upper limits to estimate the upper limit for the mass of the central black hole.

We use relations published in \citet{novak_correlations_2006} to estimate the mass of the central black hole. For the \textit{R} band data we use the relation based on data from \citet{bettoni_black_2003}. We use the same relation for the filters \textit{F814W}, \textit{F702W} and \textit{F28x50LP} as the differences between these bands are minimal and no colour-corrections for these filters are available. For the \textit{K} band data, we use colour corrections from \citet{poggianti_k_1997} to transform the magnitude to the \textit{J} band, we use the relation based on data from \citet{marconi_relation_2003} to derive the black hole mass.

Estimating the central black hole mass from the optical data gives values of $log M_{BH}\lesssim9$. The results from the only available NIR measurement gives a value of $log M_{BH}\approx10$. The big difference is caused by the fact that combining the optical and NIR results gives a colour that strongly deviates from colours of normal elliptical or disc galaxies at similar redshift. Note that relations as presented in \citet{novak_correlations_2006} only hold for galaxies with normal SEDs. If the galaxy strongly deviates from a normal SED, black hole masses derived using luminosities are no longer reliable. However, upper limits for optical host galaxy luminosities and black hole masses derived from optical data hold.

\begin{figure}
\includegraphics[width=8cm]{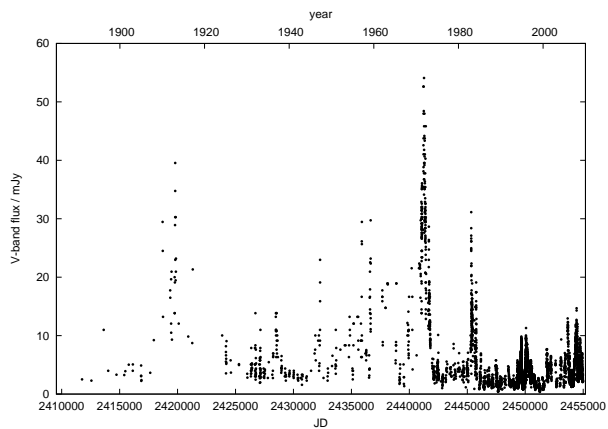}
\caption{OJ287 \textit{V}-band flux from 1885-2009. \textit{R}-band fluxes were converted to \textit{V}-band fluxes using \textit{V-R} = 0.4 mag.}
\label{historicflux}
\end{figure}

After discussing the properties of OJ287 not related to the light-curve, we will continue by putting together a list of prominent properties of the light-curve and finish by presenting a clear list of requirements which we use to rate the different models. A plot of the entire flux monitoring data, from 1885 to 2009 is presented in Fig. \ref{historicflux}.

\textbf{Exact timing} has received a lot of attention, especially in the Lehto \& Valtonen model and its successors. However, in order to be able to calculate results from their models, all authors had to make simplifications. Including all possible influences (i.e. MHD and general relativistic effects, warped disc...) would yield models with too many adjustable parameters. Additionally, solving the problem of a binary black hole together with an accretion flow would require fully relativistic MHD simulations that also calculate the radiation transfer, this is not possible with current MHD models and computer power. Therefore, we do not believe that it is possible to calculate the timing of the bursts with an accuracy in the order of days. Additionally, the light-curve has 'holes' caused by OJ287 being to close to the sun to be observed. As we discussed in Section \ref{sec:double-bursts}, this can have the effect that the exact time of the outburst can only be determined with an accuracy in the order of weeks to months. Also, before $\sim$1970 the sampling was too sparse to determine the exact time of the outbursts accurately, this effects the calculations for the models. Therefore we believe that neither are exact timing predictions possible nor is it possible to derive exact outburst times from the light-curve. Due to this fact we will not put a lot of weight on high timing accuracy in this discussion. However, we do discuss if models strongly deviate from the measured time of the outburst.

\textbf{The 'pseudo-periodicity'}: this rather awkward word describing the light-curve of OJ287 was introduced to emphasize that the outbursts in OJ287 appear regularly while deviating from periodicity. Particularly the 2005 burst strongly deviates from periodicity, being observed about a year earlier than expected for strict periodicity. It is true that strong, rather regularly appearing outbursts have been observed from the 1970s till the present. However, this represents only four cycles. We do not believe that earlier outbursts can clearly be established as the sampling was too sparse (see Fig. \ref{historicflux}). Especially as there are about 10 year long gaps in the data. The fact that outbursts were observed around the time expected before 1970 does not prove that there were periodic outbursts -- not even pseudo-periodic outbursts for that matter. For certainty, one can only say that the data before the 1970s does not contradict regularly appearing outbursts at $\sim$ 12 yr intervals. Additionally, OJ287 may be 'loosing' its regularly appearing outbursts. The peak fluxes in the double peaked outbursts have been declining significantly since the first well-sampled burst in the $\sim$1970s (see Fig. \ref{historicflux}). Between the 1994--1995 burst and the present burst there was an additional outburst which was approximately as strong as the 2005--2007 major outbursts (see Fig. \ref{historicflux}). Also, a 'third major burst' directly after the second major burst was observed during the 2005--2007 campaign. The current major bursts are by no means as exceptional as the bursts observed in the 1970s and 1980s. While bursts are still observed near the expected time, and partially more than one year earlier than expected, additional bursts of similar strength are observed. A model would thus have to explain regularly appearing outbursts every $\sim 9$ yr (in the object restframe) over $\gtrapprox 3-4$ cycles.

\textbf{Double-peaked outbursts} are observed in the 1970s, 80s and 90s-bursts. The two bursts are of similar strength and separated by $\sim1$ yr. The 2005/2007 burst might also be classified as a double-peaked burst. However, a third peak has been observed during this period, following the second peak almost with only a short delay.

\textbf{The peak intensities} in the major optical bursts have been declining from almost 60 mJy in the 1970s to only about 15 mJy in the latest bursts. Both peaks show a similar evolution (see Fig. \ref{historicflux}). If we include earlier, poorly sampled, bursts it seems like the peak intensities have been increasing from $\sim$ 1900--1970 (see Fig. \ref{historicflux}). However, as the sampling is too poor to draw clear conclusions, we will neglect the evolution before 1970.

\textbf{Radio counterparts} are observed for some optical outbursts while they are missing for others. There is no clear pattern for the occurrence of radio counterparts. During the both the 1980s and 1990s flaring, the first burst showed a radio counterpart while the second did not \citep{valtaoja_radio_2000}. Sampling for the 1970s bursts is too sparse to draw clear conclusions \citep{valtaoja_radio_2000}. During the 2005--2007 period neither of the bursts showed a radio counterpart (Merja Tornikoski \& Anne L\"{a}hteenm\"{a}ki, private communication). While we are clearly limited by small number statistics, we can state that the first burst was never observed to have a radio-counterpart, while the second burst was observed both with and without radio-counterpart. 

\textbf{Strong changes in the optical polarization core} are observed from the 1970s till today (see Fig. \ref{historicPA}). The strength of the OPC has showed both de- and increasing phases. A flip in the preferred PA of about 90\degr has been observed in the 1990s (see Fig. \ref{historicPA}). It seems that strong changes in the core polarization follow the double-peaked bursts with a delay of $\sim1$ yr. However, the connection between the double-peaked outbursts and the OPC changes is not well established due to small number statistics.

\textbf{The limitations set by the object itself}: from host galaxy data, we can set an upper limit for the central black hole mass of $M\lid10^{9} M_{\sun}$. Additionally, we know that OJ287 is a BL Lac object and thus a FR I radio galaxy. Many observations point towards the fact that BL Lac objects are fed by radiatively inefficient and thus geometrically thick accretion flows (e.g. \citealt{baum_toward_1995}, \citealt{ghisellini_fermi_2009}). 

Thus a new model would have to explain the following observations:
\begin{itemize}
\item regularly appearing outbursts in the optical light-curve with a period of $\sim 9$ yr over $\gid$ 3-4 cycles
\item double-peaked outbursts in at least three cases, both peaks have comparable strength, the current double-peaked burst might be better classified as a triple-burst
\item the peak intensities undergo strong changes, from 1970--2007 the peak intensities have been abating, both peaks show a similar evolution
\item radio counterparts are observed in some bursts, never has one of the first bursts been observed to show a radio counterpart, the second bursts are observed both with and without radio counterparts
\item the strength of the OPC changes, including a flip, changes are possibly related to the double-peaked bursts
\item a black hole mass $\lid 10^{9} M_{\sun}$
\item a radiatively inefficient, geometrically thick accretion flow
\end{itemize}

In Tab. \ref{sum_models} we present a comprehensive and short summary of the different models using our list of requirements. Each model will be discussed separately and in detail in the following paragraphs.

\begin{table*}
\begin{minipage}{180mm}
\caption{Summary of OJ287 models and their agreement with our list of requirements. For a detailed discussion, we refer to the paragraph for the respective models. Possible ratings are: yes (fulfils requirement), no (does not fulfil requirement), ? (in case of BH mass: model does not mention BH mass, otherwise: it is unclear if the model could explain the respective observation)}
\label{sum_models}
\begin{tabular}{cccccccc}
\hline \hline
 & timing accuracy & peaks changing? & radio counterparts? & OPC variability? & BH mass? & RIAF?\\
\hline
\citet{lehto_oj_1996}		& $\sim0.5$ yr error for 2005 burst	& no & ?  & no & no & no \\
\citet{valtonen_new_2007} 	& predicted 1 burst correctly (2007) 	& no & ?  & no & no & no \\
\citet{valtaoja_radio_2000}	& $\sim1$ yr error for 2005 burst 	& ?  & ?  & ?  & yes& ?  \\
\citet{villata_beaming_1998}	& none given				& yes& no & ?  & ?  & ?  \\
\citet{katz_precessing_1997}	& none given				& yes& no & no & ?  & ? \\
\hline
\end{tabular}
\end{minipage}
\end{table*}

\subsubsection{Lehto \& Valtonen: a binary black hole, two disc piercings, gravitational waves}
\label{sec:lehto-valtonen}

In this section we discuss both the original Lehto \& Valtonen model \citep{lehto_oj_1996} and the modified Lehto \& Valtonen model that was presented to fit the unexpectedly early outburst in 2005 \citep{valtonen_predictingnext_2006,valtonen_new_2007}. Note that both models are almost identical, the differences are due to the addition of a 'live disc', a new fit to the data using the 2005 burst and changed timing for earlier bursts. The basic properties of the model however are identical.

Before we compare this model to our list of requirements, we would like to discuss some properties and parameters of this model. One of the critical points in the Lehto \& Valtonen model is the question how the disturbance in the disc will affect the jet. \citet{valtonen_tidally_2009} estimated that the particles reach the jet after only a few days, producing a jet burst immediately after the first thermal flare. The jet flare merges with the accretion disc flare, no delay at all is observed between the two flares. Thus each observed burst actually consists of two flares: the accretion-disc-flare and the jet-response-flare. To achieve such a short delay, they assumed that the particles enter the jet flow as soon as they pass a radius of 10 $R_{Schwarzschild}$, which is very close to the radius at which the secondary black hole pierces the disc. Particles that get injected into the jet become visible immediately. It is very likely that changing this parameter would change the appearance of the bursts dramatically. As the short time delay between the accretion disc piercing and the jet response is so fundamentally important for this model, we will discuss this point in greater detail.

It is unclear why \citet{valtonen_tidally_2009} chose this particular radius. For comparison, the radius at which the accreting material plunges into the black hole, the innermost stable circular orbit (ISCO) is $R_{ISCO} = 3 \times R_{Schwarzschild}$ for a non-rotating black hole. The ISCO shrinks considerably if the black hole has a high spin. The ISCO is the smallest radius at which the accreting material could enter the jet. Additionally, \citet{valtonen_tidally_2009} did not take viscosity and MHD effects into account. Viscosity and turbulence play a major role in accretion processes \citep{balbus_instability_1998} and thus estimations neglecting these effects might yield wrong results.

Thus, we will estimate the time delay between the disc piercing and the jet response and compare it to the results from \citet{valtonen_tidally_2009}. If we assume that the launching of the jet happens anywhere between the disc piercing and the ISCO, we can try to estimate the time it takes the disturbance to become visible in the jet. Estimating these time-scales is rather challenging. However, we can use the viscous time scale at the radius of the disc piercing to get a lower limit for the time it takes for the disturbance to reach the ISCO. The viscous time-scale for standard $\alpha$-discs can be calculated as follows \citep{czerny_role_2006}:

\begin{equation}
 t_{visc} = 10^{7} \times \alpha^{-1}_{0.1} \times (\dfrac{r}{h_{d}})^{2} \times R_{3}^{\frac{3}{2}} \times M_{8}
\end{equation}
Where $\alpha$ is the disc viscosity, $r h_{d}^{-1}$ is the scale heights of the disc, $R_{3}$ is the radius in units of three Schwarzschild-radii and $M_{8}$ is the black hole mass in units of $10^{8} M_{\sun}$. We assume a black hole mass of $10^{10} M_{\sun}$, a disc viscosity $\alpha = 0.1$, a standard scale heights of $r h_{d}^{-1} = 10$. We estimate the radius as the semi-major axes according to the third Kepler law with the parameters given by the Valtonen model. The viscous time-scale is proportional to the black hole mass, due to the extreme black hole mass of $10^{10} M_{\sun}$ the time-scales are extremely long, the viscous time-scale and thus the lower limit for the time it takes to reach the ISCO is $\sim$ 10 yr. This is approximately three orders of magnitude longer than the delay of several days estimated by \citet{valtonen_massive_2008}. Note however that these estimates only hold for $\alpha$--discs and not for radiatively inefficient accretion flows. Simple estimates for such flows do not exist as the physics in these jets is less well known.

While we do not know at which radius the jet is launched, we can for sure say that any value considerably smaller than the orbital radius causes serious problems for the Lehto \& Valtonen model as the jet flare would be significantly delayed, possibly more than 10 years. And there might even be a longer delay as it is not clear how close to the injection point the visible jet starts. There might be a long time span between the injection of matter and the jet response.

We will now discuss the timing accuracy for the two Lehto \& Valtonen models: \citet{lehto_oj_1996} predicted the first outburst of the 2005--2007 period to happen around 2006 May 12, approximately half a year later than actually observed. In \citet{valtonen_new_2007}, the Lehto \& Valtonen model was modified, using a 'live disc' as opposed to a 'rigid disc'. They were able to explain the 'premature' 2005 flare. Thus, we can conclude that the original Lehto \& Valtonen model is not consistent with the early 2005 burst. The modified model by \citet{valtonen_new_2007} could so far only be tested during the 2007 burst and is consistent with the data.

The Lehto \& Valtonen model predicts that each outbursts starts with a very short unpolarized outburst, caused by the piercing of the secondary black hole through the accretion disc. As the later part of the outburst in this model is a jet burst, the polarization should then rise to high values. In the 2005 burst, we do not have enough polarimetric data points to assess if the data fits the model. For the 2007 burst, we indeed observe a short unpolarized flare right after the summer gap, in agreement with the Lehto \& Valtonen model. The model is also consistent with the finding that both flares show a flattening of the optical spectrum.

The Lehto \& Valtonen model has troubles explaining the observed variations in the peak intensities. The difference between the disc crossings in the Lehto \& Valtonen model is the radius at which the secondary black hole crosses the disc. The radius will probably have influence on the strength of the burst. However, while the first disc crossing moves closer to the primary black hole, the second one moves further out, thus one would expect that one of the bursts gets fainter while the other burst gets brighter. We observe a nearly simultaneous evolution of the peak intensities. Thus, we conclude that this model cannot explain the observed changes in the peak intensities. \citet{valtaoja_radio_2000} argued that the Lehto \& Valtonen model cannot explain the missing of radio counterparts in some bursts. However, in case jet induced flares without radio counterparts are possible, the Lehto \& Valtonen model would be consistent with the data. For example, an outburst in a part of the jet that is opaque for radio emission could explain bursts that do not have radio counterparts. However, it is unclear why some bursts should be opaque to radio emission while others are not. OPC variability is expected in this model, however, the model predicts movements of the jet direction at every disc crossing, while only a single very sudden movement is observed. The black hole mass of $M_{BH}=1.8\times10^{10}M_{\sun}$ assumed in this model is in clear contradiction to the estimates from host galaxy data. This black hole mass is on the very extreme end of the black hole mass distribution, especially considering the low redshift of only $z=0.306$ \citep{vestergaard_mass_2007}. To revoke this contradiction, \citet{valtonen_massive_2008} argue that OJ287 is one of the brightest AGN. But OJ287 is a blazar, the radiation from the jet is highly beamed, boosting the flux. Thus it is misleading to compare the absolute magnitude of the highly beamed blazar OJ287 to normal, mostly unbeamed AGN. Furthermore, the model explicitly assumes a thin disc, in presence of a geometrically thick disc, the bursts would lengthen considerable as the secondary black hole would remain 'inside the disc' for a considerable fraction of the orbit. Note that the disc-piercing burst in the Lehto \& Valtonen only lasts about a week, which corresponds to less than one per cent of the orbital period or an opening angle of the disc of only $\sim$1\degr.

\subsubsection{Valtaoja: a binary black hole, one disc piercing, one jet response}
\label{sec:valtaoja}

\citet{valtaoja_radio_2000} presented a model containing a binary black hole. In this model, the secondary black hole is considerably less massive than the primary and the two black holes are moving around the common centre of mass on highly eccentric orbits. Because of the high eccentricity, the secondary only pierces the accretion disc when is close to periastron, whereas the apocentre lies outside the accretion disc and thus does not cause interaction with the disc. In this model the first peak of the major outburst represents the piercing of the secondary through the accretion disc of the primary. The second peak of the outburst represents the shock front in the jet caused by the disturbance of the accretion disc.

Before we discuss if this model fulfils the requirements, we try to estimate if a gap of $\sim 1$ yr between the first and the second burst is reasonable. We use the same formula and argument as in Section \ref{sec:lehto-valtonen} to estimate this time span. We assume a standard thin disc and a black hole mass of $10^{9} M_{\sun}$. Due to the fact that the orbit is highly eccentric, we take a lower limit for the radius of 3 $R_{Schwarzschild}$. This yields a lower limit for the accretion time of $\sim 3y$. This is still in the order of magnitude of the observed time gap. Note however that this value is the absolute lower limit. On the other hand, it is not clear how that estimate would change if one assumed a radiatively inefficient instead of an standard $\alpha$-disc.

As for the timing, the Valtaoja model assumes strict periodicity in the first peak. The delay between the first and second burst is assumed to be approximately $1$ yr. However, due to the fact that the second peak represents the jet response to the disc crossing, which is obviously a very complex process, \citet{valtaoja_radio_2000} do not provide a time prediction for the second burst. Using an orbital solutions with a zero-point of $JD_{0} = 2449667$ and a period of $P = 4324 d$, the primary outburst is expected in 2006 September 25 \citep{valtaoja_radio_2000}. Thus, the primary outburst happened more than one year earlier than predicted.

In the Valtaoja model the first burst is a thermal flare, thus one would not expect high degrees of polarization. The few data points observed during the 2005 burst do show rather high degrees of polarization, in contradiction to the model. The second burst in the Valtaoja model is a jet burst and is therefore expected to be very highly polarized. Indeed, the 2007 outburst was very highly polarized. The fact that both outbursts show a flattening of the optical spectrum and thus most likely have the same origin speaks against the Valtaoja model.

As for the list of requirements: it is not evident if the Valtaoja model can explain the strong variability in the peak intensities. The first bursts is caused by a disc-crossing. Therefore considerable changes in the strength of this bursts could either be explained by changes in the radius at which the disc crossing occurs or by significant changes in the disc itself. However, if such a change in the intensity of the first burst would appear, one would also expect the second burst to reflect these changes. This is indeed observed. Thus, the model might be able to explain the changes in the peak intensities. \citet{valtaoja_radio_2000} based their models on the fact that during the double-peaked burst 1994--1995, the first peak did not have a radio-counterpart, while the second peak also appeared in radio. During the 2005--2007 burst none of the peaks had a radio counterpart. As for the Lehto \& Valtonen model, the missing of radio-counterparts could be explained if some burst happened in areas that are opaque to radio emission. Also, it is unclear if the Valtaoja model could explain the OPC variability. As for the black hole mass, \citet{valtaoja_radio_2000} based his model on a $10^{9}\times M_{\sun}$ black hole, which is the value estimated from host galaxy data. If we assume that OJ287 is powered by a radiatively inefficient accretion flow, this could pose a problem for the Valtaoja model. Even the orbit is highly eccentric, a geometrically thick disc would considerably increase the time the secondary black hole spends 'inside the disc' thereby increasing the length of the first burst. Because the exact orbit is not specified in the Valtaoja model, it is unclear how big the effect would be. Also, the limitations by the geometrically thick accretion flow are not as severe as for the Lehto \& Valtonen model as the disc-piercing-flare of the Valtaoja model lasts 0.5--1 yr as opposed to $\sim$ 1 week for the Lehto \& Valtonen model.

\subsubsection{Villata: a binary black hole, two precessing jets, beaming events}
\label{sec:villata}

\citet{villata_beaming_1998} proposed a model containing a binary black hole. Unlike in all other models, both supermassive black holes produce relativistic jets. According to this model, both major outbursts are beaming events, each of the two bursts caused by one particular jet.

Due to the fact that this is a qualitative model, no exact timing predictions are available. In this model, the peak intensities do undergo strong changes in a similar pattern as observed \citep{villata_beaming_1998}. However, as \citet{valtaoja_radio_2000} pointed out, this model cannot explain the fact that some bursts have radio counterparts while others do not. For beaming events, one would expect all wavelength ranges to show enhanced emission, but this is not observed. Due to the fact that in the Villata model the two jets 'wind around each other', building a helical structure, it is not clear how this would show in the OPC evolution. As mentioned above, the model is qualitative, thus no statement about the black hole mass is made. The Villata model does not assume any particular accretion flow, thus, it is consistent with a RIAF.

\subsubsection{Katz: a binary black hole, a precessing disc and jet}
\label{sec:katz}

\citet{katz_precessing_1997} proposed a precessing disc to be the cause for the observed behaviour in OJ 287. A precessing disc would cause the jet to precess as well. In this model, the outbursts represent a sweeping of the jet through the line of sight. The double-peaked structure of the outbursts is thought to be caused by a nodding motion of the jet precession.

As this is a qualitative model, no exact timing predictions are available. Due to the fact that both outbursts in this model are caused by beaming events from the same jet, one would expect that the strength of outbursts is variable in a pattern similar to the observations. It might even be expected that the peak luminosities undergo strong changes if the beaming factor changes between the different bursts. This could show as an additional periodic variability in the peak intensities, which would be consistent with the data. Due to the fact that both bursts are 'beaming bursts', this model cannot explain the missing of radio counterparts in some bursts, as pointed out by \citet{valtaoja_radio_2000}. It is unclear if such a setup could explain the observed OPC variations. The model is qualitative, and therefore does not specify the black hole mass. The model explicitly assumes a thin accretion disc. It is questionable that a RIAF could show similar precession as RIAFs are not rotationally supported.

\subsubsection{Alternative explanations: accretion as a cause for regular bursts or better BBH models?}
\label{sec:newmodel}

As we saw from the discussion in the previous section, none of the existing models is able to explain all observations. Several models based on the assumption of a binary black hole have been developed \citep{lehto_oj_1996,katz_precessing_1997,villata_beaming_1998,valtaoja_radio_2000} and modified (e.g. \citealt{valtonen_new_2007,valtonen_tidally_2009}). Parameters have been added, but still, none of the models can explain all the observations.

Binary black hole models might profit from growing interest in understanding binary black hole mergers. For example, it has been shown that binary black holes might show three disc systems, consisting of circum-black hole discs around the two black holes and a circum-binary disc around the whole system \citep{hayasaki_binary_2007}. However, those simulations were done for rather large separations, and it is unclear how such a system would look in a closer stage of its evolution. \citet{perego_mass_2009} studied the interaction between binary black holes and the accreting material. Considering such effects, current binary black hole models could be improved.

While gravitation based periodicity (i.e. two bodies circling around each other) is clearly the most obvious approach to explain regularity, it has been suggested that it is possible to achieve regularity through resonance in accretion discs or jets \citep{ouyed_episodic_1997}. The periodicity in simulated jets found by \citet{ouyed_episodic_1997} happened on time-scales of days and was strictly periodic and not abating. However, \citet{ouyed_episodic_1997} showed that opposed to pictures of purely chaotic jets, reappearing, resonance based events are possible in jets and accretion discs. For the case of OJ287, \citet{marscher_oj287_1997} argued that the properties of certain outbursts suggest the jet as a source for the variations. Therefore, we would like to discuss jet- and accretion-disc based approaches to explain the observed behaviour in OJ287.

We base our discussion on our finding of the optical polarization core and its evolution during the last decades. We argued in Section \ref{sec:discuss-general} that the observed behaviour in OJ287 is a sign of a gradual strengthening of the poloidal magnetic field component during the last decades. The jet magnetic field used to be dominated by a toroidal component, while it is currently dominated by a poloidal component. We also argued that due to the fact that OJ287 is a BL Lac type object it is most likely powered by a radiatively inefficient accretion flow \citep{baum_toward_1995,ghisellini_fermi_2009}. Those flows are minimally rotating \citep{igumenshchev_radiatively_2004} and therefore make it possible that accreted poloidal field lines are only minimally spun up and therefore preserve their poloidal component.

We argue that it would be rather astonishing to observe two exceptional processes, namely regular bursts and changes in the jet magnetic field, going on in one object without any connection between the two. We therefore suggest a casual connection between the regular bursts and the OPC variability, with the OPC variability being the 'main symptom' and the regular bursts being a 'after-effect' of the changes in the jet. 

We suggest that the bursts are related to the accretion of magnetic field lines. We suggest that the regularly appearing flares are signs that the accretion of the magnetic field happened in avalanches. Let us assume that at some point in the past, massive accretion of magnetic field caused strong disturbances in the magnetic field of the accretion disc. This disturbance caused a resonance in the accretion disc, a 'magnetic breathing' of the disc. The resonance causes regularly appearing avalanche accretion of magnetic field. We suggest that each double-peaked flare represents a phase of massive magnetic field accretion. As the central (poloidal) magnetic field strengthens, it starts to dominate the area near the black hole. The strong central magnetic field starts to impede further accretion of magnetic field of the same orientation. Due to this, the double-peaked flares are weakening over the course of time. Finally, the strong central magnetic field will 'choke' the resonance.

Based on this assumption, we interpret that the first episode of avalanche accretion happened prior to the 1970s bursts. This event caused the magnetic breathing. Phases of avalanche accretion were observed in the early 1970s, 1984--1985, 1994--1995 and 2005--2007, showing as gigantic double-peaked flares. In our interpretation, the toroidal magnetic field component fully dominated the jet during the early phase of photopolarimetric monitoring. This showed in a strong alignment of the OPC along the jet direction between the 1970s and 1980s bursts. The avalanche accretion during the 1980s bursts caused a strengthening of the poloidal magnetic field component. The poloidal and toroidal component were of similar strength and thus the OPC was rather weak. The next accretion during the big flaring around 1994 further strengthened the poloidal component and it became dominant, causing the OPC to align perpendicular to the jet. We interpret the abating of the peak intensities in the double-peaked flares as a sign of the strengthening of the central poloidal magnetic field. The central field starts to repel further accretion of poloidal magnetic field lines and therefore the strength of the double-peaked bursts weakens. According to our interpretation, the fact that the periodicity in OJ287 is lost and the strength of the double-peaked bursts has reached the level of ordinary outbursts can be interpreted such that the central poloidal magnetic field has become so strong that the avalanche accretion of magnetic field lines of the same orientation is no longer possible. The resonance got choked by the strong central field. Therefore, we predict that no more regularly appearing double-peaked outbursts will be observed in the future. Based on our assumptions, these burst were caused by resonant avalanche accretion which is nowadays impeded by the strength of the central magnetic field. This explanation is also well consistent with our finding that the OPC strengthened after the 2007 burst.

If this interpretation were true, the OPC in OJ287 would remain strong and oriented perpendicular to the jet direction. Due to the fact that no more accretion of open magnetic field lines of considerable strength is expected, the jet magnetic field, and thus the OPC, should be stable and not show strong changes. Also, no further strong double-peaked bursts are expected based on this assumption as the central field has become so strong that avalanche accretion is simply not possible any longer.  

We are well aware that the suggested explanation is speculative. While it has been shown in simulations that setups as the one described in the 'magnetic breathing' model are realistic (see e.g. \citealt{igumenshchev_three-dimensional_2003}), some points are not fully clear. It is not clear if a timespan of $\sim$10 yr is realistic for such a resonance behaviour. As argued earlier, time-scales of years are normal for accretion processes around supermassive black holes, thus we believe the observed time-scale is realistic for resonance processes in the accretion disc. The biggest caveat of this approach is the fact that it does not naturally explain the double-peaked bursts. We argue that what we see is the accretion of magnetic field coupled to accreting matter. While the first bursts represents the accretion of the magnetic field, which can be observed instantaneously, the matter needs time to reach the optical jet. The delay of about one year between the first and the second burst in this model would represent the time it takes the matter to reach the visible jet. This would explain that both bursts show a similar evolution in strength. However, it does not explain the missing of radio-counterparts in some bursts. It is not yet clear what decides if a disturbance in the jet will be observed in radio, therefore it is unclear if the magnetic breathing model can explain the radio behaviour. The 'magnetic breathing' approach does explain the OPC variability, it is consistent with a moderate black hole mass and a radiatively inefficient accretion flow.

With more advanced simulations, it will hopefully soon become possible to simulate both accretion processes and the launching of the jet simultaneously. When those simulations become possible, it will be interesting to assess if a setup as suggested in this section is realistic. It will also be interesting to see if developments in the theoretical understandings of jets will bring new explanations. Further observations will also have to show if the strong regularly appearing outburst have indeed stopped. Till then, the case of OJ287 will remain a mystery.

\section{SUMMARY AND CONCLUSIONS}
\label{sec:conclusions}

In this paper we present photopolarimetric monitoring of OJ287 during the most recent major double-peaked outburst. We analyse the photopolarimetric light-curve, we describe the general appearance of the light-curve, analyse statistical properties and study the appearance of flares in polarization. We also compose a list of requirements for OJ287 models in which we also include properties of OJ287 not related to the light-curve. We use this list to compare all proposed models. We also discuss alternative explanations for OJ287. Our finding can be summarized as follows:

\begin{itemize}
\item During the 2005--2009 monitoring campaign, we observe two major outbursts. The first burst starts in summer 2005 and lasts till early summer 2006. The second burst starts in early summer 2007, it reaches its peak between mid August and 2007 September 12. The second burst is considerably longer lasting than the first one, covering an overall timespan of almost 2 years, as opposed to about 1 yr in the first burst. Both burst show strong polarization in both the maximum and the declining phase.
\item The degree of polarization does not correlate with the flux, in agreement with findings of other authors \citep{jannuzi_optical_1994}.
\item We find a strong preferred position angle, caused by an underlying stable source of polarized emission: the optical polarization core (OPC). The OPC is stable on time scales, however, we did observe a fast change in the OPC during our monitoring campaign. The change happened after the 2007 burst and strengthened the OPC considerably. We also found changes of the OPC during the last decades, showing as a migration in the Stokes plane crossing the zero point. Those changes are also correlated with the double-peaked bursts. We interpret that the OPC emission originates from the quiescent jet. Our observations indicate that a global magnetic field exists in blazar jets and can be observed as alignment of the optical polarization position angle. Based on former studies on the alignment of the optical polarization in blazars \citep{angel_optical_1980,jannuzi_optical_1994}, we find a difference between the jets of FSRQs and BL Lacs. In BL Lacs, the optical polarized emission is dominated by the quiescent jet, while in FSRQs it is dominated by the turbulent emission. Further observations will have to show if this holds for bigger samples.
\item We identify five events in the polarized emission that we classify as 'Bubbles' and two events we classify as 'Swings'. Both Bubbles and Swings are circular movements in the Stokes plane, generally known as rotators. No common centre for all rotators can be identified, this cannot be explained through fast OPC variability. All rotators happen on similar time-scales and show a preferred direction of rotation. We interpret these events as movements of shock-fronts both up- and downstream the jet. Through their movement, the shock fronts swipe through different orientation of the helical magnetic field, thereby producing swinging motions.
\item We study optical spectral index variations and find a 'bluer-when-brighter' trend in both flares. This is in agreement with other studies of OJ287 \citep{hagen-thorn_variable_1998}.
\item We compose a list of requirements for OJ287 models. The list includes: regularly appearing double peaked outbursts, strong variations in the peak intensities, differences between the bursts regarding the existence of radio counterparts, strong changes in the preferred position angle (which we interpret as variability in the OPC), a black hole mass of $M_{BH} \lesssim 10^{9} M_{\sun}$ and a radiatively inefficient, geometrically thick accretion flow.
\item We assess all models using the list of requirements. We conclude that none of the models can explain all items on the list. The original Lehto \& Valtonen model \citep{lehto_oj_1996} did not predict the early 2005 outburst. Its successor, the new Lehto \& Valtonen model \citep{valtonen_predictingnext_2006,valtonen_new_2007} is consistent with the timing of the 2007 burst. However, it cannot explain any of the other items on our list and is in contradiction with four items on the list. The Valtaoja model \citep{valtaoja_radio_2000} did not predict the early 2005 outburst, no modification to the model were made to fit the new observations. It is not clear if the Valtaoja model can explain the other items on the list. Both the Villata model \citep{villata_beaming_1998} and the Katz model \citep{katz_precessing_1997} cannot explain all items on the list.
\item We also discuss possibilities for alternative explanations. We discuss a new approach based on the finding of violent variability of the optical core polarization in OJ287, observed since the 1970s. We argue that the optical polarization core is a sign of the quiescent jet and thus traces the jet magnetic field. Thus a change in the direction of the optical polarization core indicates a change of the direction of the jet magnetic field. We suggest that this could have taken place in OJ287 through accretion of poloidal magnetic field lines. We suggest that a strong accretion event in the past caused resonance in the accretion disc. This resonance showed as avalanche accretion of magnetic field lines every nine years in the object restframe, causing the double-peaked burst. According to our interpretation, the strength of the double-peaked outbursts is abating as the central magnetic field strengthens and thereby repels accreting magnetic field lines. We argue that the central magnetic field started to choke the resonance and therefore we expect no more regularly appearing bursts. We also conclude that as long as this toy model cannot be confirmed by simulations, the mystery of OJ287 will remain unsolved.
\end{itemize}

\section*{ACKNOWLEDGEMENTS}
The authors would like to thank the anonymous referee for his comments and suggestions. We would also like to thank Staszek Zola, Merja Tornikoski, Anne L\"{a}hteenm\"{a}ki, the Mets\"{a}hovi monitoring team, Esko Valtaoja and Vilppu Piirola. Part of the data presented here are based on observations made with the Nordic Optical Telescope, operated on the island of La Palma jointly by Denmark, Finland, Iceland, Norway, and Sweden, in the Spanish Observatorio del Roque de los Muchachos of the Instituto de Astroﬁsica de Canarias. Part of the data presented here have been taken using ALFOSC, which is owned by the Instituto de Astrofisica de Andalucia (IAA) and operated at the Nordic Optical Telescope under agreement between IAA and the NBIfAFG of the Astronomical Observatory of Copenhagen. Based on observations collected at the Centro Astron\'{o}mico Hispano Alem\'{a}n (CAHA) at Calar Alto, operated jointly by the Max-Planck Institut f\"{u}r Astronomie and the Instituto de Astrof\'{\i}sica de Andaluc\'{\i}a (CSIC). This work was partly supported by the Polish MNiSW grant No. 3812/B/H03/2009/36. Jianghua Wu and Xu Zhou are supported by the Chinese National Natural Science Foundation grants 10603006 and 10633020.

\bibliographystyle{mn2e}
\bibliography{OJ.bib}

\label{lastpage}
\end{document}